\documentclass[aps,prd,reprint,amsmath,noeprint]{revtex4-2}
\usepackage{orcidlink}
\usepackage{multirow}
\allowdisplaybreaks[1]

\begin{document}
\title{Phenomenological analysis of charmless $B\to PV$ decays \\ in the modified perturbative QCD approach}
\author{Sheng L\"u\,\orcidlink{0000-0002-5266-8024}} \email{shenglyu@mail.nankai.edu.cn}
\author{Ru-Xuan Wang\,\orcidlink{0000-0002-6394-1225}} \email{wangrx@mail.nankai.edu.cn}
\author{Mao-Zhi Yang\,\orcidlink{0000-0002-5052-713X}} \email{yangmz@nankai.edu.cn}
\affiliation{School of Physics, Nankai University, Tianjin 300071, People's Republic of China}
\date{\today}

\begin{abstract}
	We analyze the charmless two-body decays of the $B$ meson into a pseudoscalar and a vector meson, referred to as $B\to PV$ decay, within the framework of the modified perturbative QCD (PQCD) approach, which was originally developed to solve the long-standing puzzles, the so-called $\pi\pi$ and $\pi K$ puzzles. In this method, based on the conventional PQCD calculations, soft form factors and contributions arising from color-octet quark-antiquark components in the final state are introduced, which involve essentially nonperturbative dynamics. By employing the flavor SU(3) symmetry and its breaking effect, the parameters describing soft contributions are correlated and subsequently reduced into a more concise set of SU(3) parameters. Through a $\chi^2$ analysis procedure applied to the experimental data, these parameters are determined, from which the corresponding theoretical results for the branching ratios and $CP$ asymmetries in $B\to PV$ decays are derived. Within this framework, it is demonstrated that the theoretical calculations align well with experimental measurements, suggesting that the method developed for solving $\pi\pi$ and $\pi K$ puzzles can be extended to more decay modes of the $B$ meson. The numerical analysis shows that these soft contributions have a significant influence on the theoretical results. Furthermore, more precise experimental data are needed for deeper investigations of the $CP$ asymmetries.
\end{abstract}

\maketitle

\section{Introduction \label{sec:intro}}
The study of the hadronic weak decay of the $B$ meson is essential for the rigorous validation of the Standard Model (SM). It is helpful in understanding the complex mechanics of the strong interaction, revealing the origin of $CP$ asymmetries, and providing invaluable insights into potential signals of new physics \cite{BaBar:2001pki,Belle:2001zzw,HFLAV:2022esi}. In this work, we concentrate on the charmless two-body $B\to PV$ decays, where $P$ and $V$ represent the light pseudoscalar and vector mesons, respectively. In recent decades, with the upgrade and operation of the Large Hadron Collider (LHC) and the Super-KEKB, the LHCb and Belle-II collaborations have collected a huge amount of data. The measurements for most $B\to PV$ decays have been refined, achieving a level of precision that is unprecedented \cite{Belle-II:2018jsg,LHCb:2017pkd,LHCb:2019xmb,LHCb:2019jta,LHCb:2020wrt,LHCb:2022tuk,Belle:2022dgi,Belle-II:2023uql}. Consequently, to fully make use of the experimental data, higher-precision theoretical calculations has become increasingly necessary. 

Due to the intricate nonperturbative dynamics inherent in QCD, the theoretical calculation of observables in hadronic $B$ meson decay presents significant challenges. Conventionally, by employing certain factorization hypotheses, the matrix elements for hadronic two-body $B$ decays can be expressed as a product of two components. One component denotes the transition of the heavy $b$ quark into light quarks, while the other accounts for the hadronization of the final quarks. Based on these hypotheses, various factorization approaches have been developed, including the QCD factorization (QCDF) \cite{Beneke:1999br,Beneke:2000ry,Beneke:2001ev,Beneke:2003zv}, perturbative QCD (PQCD) \cite{Keum:2000ph,Keum:2000wi,Keum:2000ms,Lu:2000em,Lu:2000hj}, and soft collinear effective theory (SCET) \cite{Bauer:2000ew,Bauer:2000yr,Bauer:2001ct}. Additionally, there are approaches originating from general principles, such as SU(3) flavor symmetry and topological properties of diagrams \cite{Berthiaume:2023kmp,Bhattacharya:2025wcq,BurgosMarcos:2025xja,Fang:2026fhl}. These approaches, like the topological diagram approach (TDA) and the factorization-assisted topological-amplitude approach (FAT), aim to extract the QCD effects from experimental data and implement model-independent analyses of SU(3) data fitting \cite{Chiang:2006ih,Chiang:2008zb,Cheng:2014rfa,Zhou:2016jkv,Cheng:2022ysn}. Each of these methods possesses its distinctive advantages and limitations, offering valuable insights into the experimental results within their respective domains of applicability. However, several long-standing puzzles remain to be resolved, particularly concerning the puzzles associated with the $\pi\pi$, $K\pi$, $\pi\rho$, $K\rho$, and $\pi K^*$ decay channels. These puzzles are: 1) On the one hand, the predicted branching ratio of $B^0\to \pi^0\pi^0$ decay is several times smaller than the experimental measurement, on the other hand, the theoretical calculations of the branching ratios of $B^+\to\pi^+\pi^0$ and $B^0\to \pi^+\pi^-$ are consistent with experimental data \cite{Beneke:2001ev,Beneke:2003zv,Lu:2000em,keum:2002vi}. 2) The $CP$ asymmetries of $B^\pm\to \pi^0 K^\pm$ and $B^0\to \pi^\mp K^\pm$ should be approximately equal to each other according to the theoretical analysis based on the factorization theorem \cite{Li:2005kt}. However, the measured results from experiments for the $CP$ asymmetries of these two channels are dramatically different. In addition, the measured branching ratios of $B\to K\pi$ decays are also puzzling according to the theoretical analysis \cite{Buras:2003yc,Kim:2005jp,Fleischer:2007mq}. 3) The theoretical predictions to the branching ratios of $B^0\to \pi^0\rho^0$ and $\pi^0K^{*0}$ in PQCD are apparently smaller than experimental data \cite{Lu:2000hj,Chai:2022ptk}. In pursuit of resolving these puzzles, numerous efforts have been undertaken, including calculations involving higher-order and higher-power terms \cite{Li:2005kt,Wang:2008rk,Rui:2011dr,Bai:2013tsa,Zhang:2014bsa,Liu:2015upa,Chai:2022ptk,Cheng:2022ysn}.
While these attempts have narrowed the gap between experimental measurements and theoretical predictions, it remains inspiring to reexamine these puzzles from a new perspective.

To solve the $\pi\pi$ and $K\pi$ puzzles, we make several improvements to the conventional PQCD approach, as presented in Refs.\cite{Lu:2022hbp,Wang:2022ihx}. The improvements include: 1) we use a new wave function for the $B$ meson, which is the result of solving the wave equation in the QCD-inspired relativistic potential model \cite{Yang:2011ie,Liu:2013maa,Liu:2015lka,Sun:2016avp,Sun:2019xyw}, where the mass spectrum of both the $B$ meson and its higher excited states can be obtained in consistence with experimental data simultaneously. 2) To describe the residual long-distance contributions in PQCD calculations, soft form factors and an infrared cutoff scale $\mu_c$ are introduced. Contributions with the energy scales below the cutoff scale $\mu_c$ are viewed as nonperturbative and are absorbed into the soft parameters. Only the contributions above the scale $\mu_c$ are calculated by perturbative method. 3) Color-octet contributions are introduced, which arise from the interactions between the color-octet quark-antiquark pairs produced in $B$ decays after hard interactions. As the distance between the color-octet pairs increases, soft interactions emerge to change the color-octet quark-antiquark pairs into color-singlet mesons, which are described by introducing a few soft parameters. The values of these soft parameters are determined by fitting the experimental data. This framework provides a good interpretation of the experimental observables in $\pi\pi$ and $K\pi$ decays. In extending the works of \cite{Lu:2022hbp,Wang:2022ihx} to more charmless $B\to PP$ decays, where $P$ denotes a light pseudoscalar meson, we systematically incorporate the flavor SU(3) symmetry and its breaking effect \cite{Lu:2024jzn}. By virtue of this symmetry, the number of free parameters is significantly reduced, thereby enhancing the predictive power of the theory.


In this work, we advance the application of the established theoretical framework to the $B\to PV$ decays.
Through a comprehensive analysis of 32 decay channels, involving 62 experimental observables including branching ratios and $CP$ asymmetries, the values for nonperturbative parameters are obtained. The experimental data for the $\pi\rho$, $K\rho$, $\pi K^*$, and other $B\to PV$ decay channels are explained within a unified framework. In particular, the branching ratios manifest a remarkable agreement between the experimental measurements and our theoretical output, which indicates the success of our framework in accommodating the experimental data. After a detailed numerical analysis, we find that the nonperturbative contributions play an essential role for theoretically explaining the experimental data. However, notable discrepancies are observed in the $CP$ asymmetries for several decay channels, including $\pi^0\rho^0$, $\eta^\prime\rho^+$, $\pi^0 K^{*+}$ and $\eta^\prime K^{*+}$. Considering the large uncertainties of the present measurements, the further investigations require more precise experimental data for $CP$ asymmetries.

The paper is organized as follows. Sec.~\ref{sec:lo} is dedicated to explaining the theoretical framework, with a particular focus on the effective Hamiltonian and the PQCD factorization formulation. The details of our theoretical calculation are presented in Sec.~\ref{sec:cal}. Here, we perform the calculations that include the leading-order (LO) diagrams, the next-to-leading-order (NLO) corrections, and the contributions from the nonperturbative soft form factors as well as the color-octet contributions. Employing these theoretical formulations, the specific numerical analysis is expounded in Sec.~\ref{sec:num}, where we present our results for the branching ratios and $CP$ violations, comparing them with the experimental data and the previous studies, and offering a thorough discussion of the related $B\to PV$ decay channels. Finally, a brief summary is provided in Sec.~\ref{sec:sum}.

\section{The Theoretical Framework for PQCD Calculations \label{sec:lo}}
\subsection{The effective Hamiltonian \label{subsec:heff}}
The nonleptonic decay of the $B$ meson primarily occurs through the weak interaction described by exchanging the $W$ boson. For the case of $B$ meson decays, effective field theory proves to be especially useful. By integrating out the $W$ boson field in the generating functional of the Green functions, the heavy degrees of freedom are removed, transforming the full theory into an effective theory with only $b$ and lighter quark fields. Through the operator product expansion, the nonlocal action is represented by a series of local terms, with $1/m_W$ (where $m_W$ denotes the mass of the $W$ boson) serving as the expansion parameter \cite{Wilson:1969zs, Wilson:1970ag}.

The effective Hamiltonian for the nonleptonic $B$ meson weak decays has been derived and given by \cite{Buchalla:1995vs}
\begin{align}\label{eq:heff}
	\mathcal{H}_{\textrm{eff}}^{b\to q}& = \dfrac{G_F}{\sqrt{2}}\Biggl\{V_{ub}V_{uq}^*\left[C_1(\mu)O_1^u(\mu) + C_2(\mu)O_2^u(\mu)\right] \nonumber \\
		&\quad-V_{tb}V_{tq}^*\left[\sum_{i=3}^{10}C_i(\mu)O_i(\mu)+C_{8g}(\mu)O_{8g}(\mu)\right]\Biggr\},
\end{align}
where $G_F$ is the Fermi constant and $b\to q$ refers to the transitions of a $b$ quark into either a $d$ or $s$ quark. The $V_{ub(q)}^{(*)}$ and $V_{tb(q)}^{(*)}$ represent the Cabibbo-Kobayashi-Maskawa (CKM) matrix elements which characterize the weak interactions. In this effective Hamiltonian, the Wilson coefficients $C_j(\mu)$ (for $j=1-10,8g$) and the four-quark operators $O_j(\mu)$ absorb the short- and long-distance effects, respectively.
The values of the Wilson coefficients and the expressions of the effective operators can be found in  Ref. \cite{Buchalla:1995vs}.

Considering the asymptotic freedom of the QCD interaction, the strong interaction coupling constant $g_s$ becomes relatively weak at the high energy scales, making perturbation theory applicable. Thus, the Wilson coefficients, which include the short-distance effects, can be systematically derived to the desired order in the perturbative expansion, where their associated uncertainties are kept under control. However, the long-distance QCD effects remain quite intricate, constituting the principal source of uncertainty in the theoretical predictions.
In the process of the nonleptonic $B$ meson decays, the light quarks generated in the interactions of short distance are finally confined into the color-neutral hadrons. The hadronization of the light quarks is dominated by complex nonperturbative QCD dynamics. These contributions are contained within the hadronic matrix elements of the effective four-quark operators, which are challenging to calculate straightforwardly and are typically estimated with the help of various nonperturbative techniques, including factorization approaches \cite{Beneke:1999br}, lattice QCD (LQCD) \cite{Wilson:1974sk}, QCD sum rules \cite{Shifman:1978bx}, and other phenomenological models.

\subsection{The PQCD factorization approach \label{subsec:pqcdf}}
In this work, we focus on the nonleptonic charmless decays of $B$ meson with final states that consist of a pseudoscalar and a vector meson. The masses of the final-state pseudoscalar and vector mesons are significantly lighter than that of the $B$ meson. Consequently, the momenta of the outgoing final-state mesons are expected to be large, nearly lightlike, and oppositely aligned in the rest frame of the initial $B$ meson. In order to increase the spectator quark's momentum from its characteristic scale $\mathcal{O}(\Lambda_{\text{QCD}})$ in a stationary $B$ meson to the much larger scale $\mathcal{O}(m_B)$ exhibited by the outgoing meson, the QCD interactions between the $b$ quark and the spectator quark should be dominated by hard scattering in the decay process.
In this situation, the PQCD approach based on the $k_T$ factorization is particularly well suited for calculating the hadronic matrix elements \cite{Li:2003yj}.

\begin{figure*}
	\includegraphics[width=0.8\textwidth]{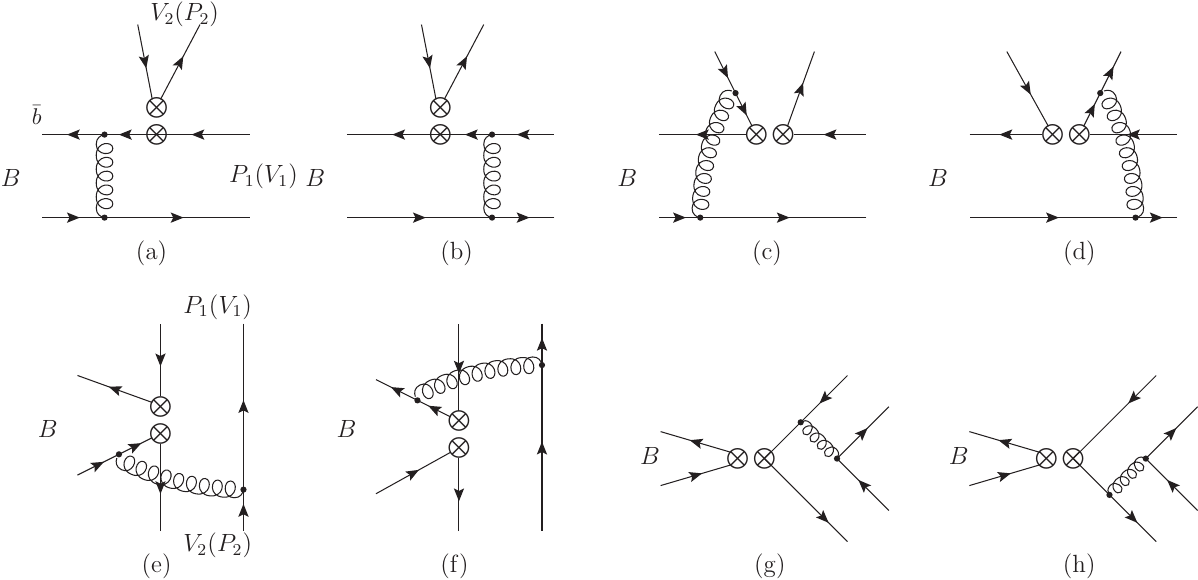}
	\caption{\label{fig:eightdiagram} Topological diagrams that contribute to the $B\to PV$ decays.
        The diagrams are grouped into four distinct categories: factorizable emission diagrams (a) and (b), nonfactorizable emission diagrams (c) and (d), nonfactorizable annihilation diagrams (e) and (f), and factorizable annihilation diagrams (g) and (h).}
\end{figure*}

The decay amplitudes of $B\to PV$ decays involve several topological diagrams, including factorizable emission, nonfactorizable emission, factorizable annihilation, and nonfactorizable annihilation, as shown in Fig.~\ref{fig:eightdiagram}.
The subscripts $1$ and $2$ are used to denote the position of the final-state mesons.

The parton momentum fractions of the constituent quarks in three mesons are denoted as $x_{0,1,2}$ respectively.
The momentum fraction $x$ must be integrated over the range between $0$ and $1$ in the exclusive process.
In the end-point region where $x$ approaches $0$, infrared divergence may emerge. The potential presence of the end-point divergence in the formula must be carefully examined. We take the factorizable emission diagram (a) in Fig.~\ref{fig:eightdiagram} as an example, since this kind of diagram provides the leading contribution in general. Within the collinear factorization framework, the lowest-order hard kernel corresponding to this diagram is proportional to $1/(x_0x_2^2)$. Taking the pion as an illustrative example, employing the asymptotic model for its distribution function, we have $\phi_\pi \propto x_2(1-x_2)$ \cite{Szczepaniak:1990dt}. The decay amplitude is represented by the convolution of the hard scattering kernel with the meson distribution amplitudes. It is apparent that the integral over $x_2$ is logarithmically divergent in the end-point region. This may imply that the process is governed by soft QCD dynamics or that the collinear factorization breaks down under these conditions. Actually, in the end-point region, the transverse momenta of the quarks, usually omitted in comparison to the dominant longitudinal components, must be retained. Incorporating the transverse momenta $\mathbf{k}_{0(1,2)T}$, the lowest-order hard kernel is modified to become proportional to $1/\{[x_0x_2m_B^2+(\mathbf{k}_{0T}-\mathbf{k}_{2T})^2](x_2m_B^2+\mathbf{k}_{2T}^2)\}$, and the decay amplitudes do not develop a divergence in the end-point region either. Hence, by employing the $k_T$ factorization, the end-point singularity is avoided.

To complete the $k_T$ factorization procedure, besides investigating the ultraviolet divergence through the renormalization method, it is necessary to isolate the infrared divergence as well. The infrared divergences arising from the radiative corrections reflect the nonperturbative property in the perturbative calculations.
In the nonleptonic decays of the $B$ meson, such divergences can be incorporated into the mesons' wave functions, as demonstrated in Refs.~\cite{Li:1994cka, Keum:2000wi}. The remaining finite part, referred to as the hard kernel, can be handled perturbatively through the Feynman diagram expansion.

The $B\to PV$ decays involve several energy scales, including the electroweak scale $m_W$, the typical scattering scale $t$, and the transverse extension scale $1/b$. The significant gaps between these scales may give rise to the large logarithmic terms, like $\log(tb)$ and $\log^2(Pb)$, where $P$ denotes the dominant light-cone momenta of mesons. In order to preserve the reliability of the perturbative expansion, these logarithms must be systematically organized. It is well established that the single logarithms can be summed through the renormalization group method, while the double logarithms must be treated using the resummation technique \cite{Li:1996rv}. The single logarithm $\log(m_W/t)$ is summed to all order and included in the Wilson coefficients $C(t)$, corresponding to the evolution from the $m_W$ down to the typical scale $t$.
And the logarithm $\log(tb)$ is summed into an exponential factor, which describes the evolution from the scale $t$ to $1/b$. The double logarithm $\log^2(Pb)$ emerges from the overlap of the soft and the collinear logarithms.
Its resummation leads to the Sudakov factor $\exp[-s(P,b)]$, which suppresses the decay amplitudes in the large $b$ region \cite{Li:1992nu}. Additionally, the logarithmic term $\log^2x$ is resummed to yield the threshold resummation factor $S_t(x)$ \cite{Li:1998is}.

The soft contribution below the scale $1/b$ is included in the meson's wave function $\Phi$. In light of its nonperturbative characteristics, $\Phi$ must be obtained by fitting the experimental data or through nonperturbative methods, such as LQCD, light-cone sum rules (LCSR), potential model, etc. On the other hand, the hard kernel, denoted as $H$, can be perturbatively calculated. Using the technique mentioned above to guarantee the self-consistency of the perturbation calculation, the factorization formula for the decay amplitude can be expressed as follows

\begin{align} \label{eq:fac}
	\mathcal{M} &= C(t) \otimes H(t,b) \otimes \Phi(x,b) \nonumber \\
		&\quad \otimes \exp\left[-s(P,b)-2\int_{1/b}^t \dfrac{d\bar{\mu}}{\bar{\mu}}\gamma_\Phi(\alpha_s(\bar{\mu}))\right],
\end{align}
where $\gamma_\Phi$ indicates the anomalous dimension of the wave function $\Phi$, and the symbol $\otimes$ emphasizes the convolution operation.

\section{The PQCD calculations and Nonperturbative Corrections \label{sec:cal}}
\subsection{The leading order PQCD amplitudes \label{subsec:loamp}}
As we have explained, the wave functions of the mesons cannot be determined within the PQCD framework.
Instead, they are treated as general inputs, independent on the specific decay channel. For the $B$ meson, we employ the wave function obtained from the QCD-inspired relativistic potential model \cite{Yang:2011ie}.
This method has been demonstrated to be able to effectively describe both the ground state and the excitation spectrum of the $B$ meson \cite{Liu:2013maa, Liu:2015lka}. Its explicit form and associated parameters are organized in the Refs.~\cite{Sun:2016avp, Sun:2019xyw}, and the summarized formulas can be also found in Ref.\cite{Wang:2024xci}.
For brevity, the further details are provided in Appendix~\ref{app:bwavefunc}.

In regard to the light mesons, $\pi$, $K$, $\eta^{(\prime)}$, $\rho$, $K^*$, $\omega$, and $\phi$, their distribution amplitudes are defined by the matrix elements on the light cone, expended up to twist-3 accuracy.
In this analysis, we neglect the contributions from the suppressed three-particle light-cone distribution amplitudes (LCDAs). Using the conformal expansion technique, these LCDAs are represented as a series in terms of the Gegenbauer polynomials. Detailed expansion of the matrix element, as well as the complete formulas for the LCDAs, is thoroughly discussed in Refs.~\cite{Braun:1989iv, Ball:2006wn, Ball:2007rt}. Additionally, elaborate expressions for the LCDAs of light pseudoscalar and vector mesons are provided in our recent works \cite{Gui:2024nlk,Wang:2024xci,Lu:2025cop}.

The decay amplitudes related to the four types of topological diagrams shown in Fig.~\ref{fig:eightdiagram} are denoted as $F_e$, $\mathcal{M}_e$, $F_a$, and $\mathcal{M}_a$ respectively, where $F_e$ denotes the decay amplitude of the factorizable emission diagrams of Figs.~\ref{fig:eightdiagram} (a) and (b), $\mathcal{M}_e$ the decay amplitude of the nonfactorizable emission diagrams of Figs.~\ref{fig:eightdiagram} (c) and (d), $F_a$ the decay amplitude of the factorizable annihilation diagrams of Figs.~\ref{fig:eightdiagram} (g) and (h), and $\mathcal{M}_a$ the amplitude of the nonfactorizable annihilation diagrams of Figs.~\ref{fig:eightdiagram} (e) and (f). Here, let's consider the decay amplitude of the $B\to P_1 V_2$ decay for the factorizable emission diagrams as an example, where the subscripts 1 and 2 denote the mesons shown in the relevant positions in Figs.~\ref{fig:eightdiagram} (a) and/or (e). The decay amplitude for the diagrams of Figs.~\ref{fig:eightdiagram}(a) and (b) with the insertion of the $(V-A)(V-A)$ four-quark operator can be written as
\begin{align} \label{eq:fe}
	&F_{e,P_1V_2} \nonumber \\
	&\; =\textstyle -4\sqrt{2N_c}\pi^2(C_F/N_c)f_B{f_{V_2}^{\parallel}}m_B^2
		\int_0^{\frac{m_B}{2}}k_{0\perp}dk_{0\perp}\int_{x_0^d}^{x_0^u}dx_0 \nonumber \\
	&\quad\times\textstyle\int_0^1dx_1 \int_0^\infty b_0db_0b_1db_1
		G_B(x_0,b_0,k_{0\perp}) \{\alpha_s(t_e) \nonumber \\
	&\quad \times [((2-x_1)E_q-x_1k_0^z) \phi_{P_1}^A(x_1,b_1)
		-r_{P_1}((1-2x_1)E_q \nonumber \\ &\quad -k_0^z) (\phi_{P_1}^P(x_1,b_1)-\phi_{P_1}^T(x_1,b_1))]
		h_e(x_0,1-x_1,b_0,b_1) \nonumber \\ &\quad \times  S_t(x_1)\exp[-S_B(t_e)-S_{P_1}(t_e)]
		+\alpha_s(t_e^\prime)2r_{P_1} \nonumber \\ &\quad \times (E_q+k_0^z)\phi_{P_1}^P(x_1,b_1)h_e(1-x_1,x_0,b_1,b_0) S_t(x_0) \nonumber \\
	&\quad \times \exp[-S_B(t_e^\prime)-S_{P_1}(t_e^\prime)]\},
\end{align}
where $\phi_{P_1}^{A,P,T}$ are the twist-2 and -3 LCDAs of the pseudoscalar meson, respectively. The function $G_B$ in the amplitudes are quantities related to the $B$-meson wave function, which is defined as
\begin{align}
	&G_B(x,b,k) \nonumber \\
	&\quad= \left(\frac{m_B}{2}+\frac{k^2}{2x^2m_B}\right) K(x,k) (E_Q+m_Q) J_0(kb).
\end{align}
The quantity $K(x,k)$ is a function related to the wave function of $B$ meson, which can be seen in Appendix \ref{app:bwavefunc}. To ensure the heavy-quark mass is real, the momentum fraction $x_0$ is constrained to lie within $x_0^{u,d}=1/2 \pm \sqrt{1/4-|k_{0\perp}|^2/m_B^2}$. $f_B$, $f_{P_1}$, and $f_{V_2}^{\parallel(\perp)}$ are the decay constants of the  mesons of $B$, $P_1$, and $V_2$, respectively. The dimensionless chiral parameter $r_{P_1}$ is defined as $r_{P_1} = m_{P_1}^2/[(m_{q}+m_{q^\prime})m_B]$, where $m_{P_1}$ is the mass of the light meson, $m_q$ and $m_{q^\prime}$ the masses of the light quarks in the light meson, and $m_B$ the mass of $B$ meson. $J_0$ is the Bessel function, which emerges from the Fourier transformation of the hard kernel from momentum space to coordinate space. The hard function $h_e$ can be found in Appendix~\ref{app:fullamp}. The threshold resummation factor $S_t$ and the Sudakov factor $S_{B(P_1)}$ have already been discussed in Sec.~\ref{subsec:pqcdf}. In order to suppress the high-order corrections, the scales $t_{e}^{1,2}$ are chosen to be the largest virtuality present in the decay process. For simplicity, the amplitudes for other topological diagrams are summarized in Appendix~\ref{app:fullamp}, where the detailed expressions for $t_e$ and $h_e$ are included.

\subsection{The next-to-leading-order corrections \label{subsec:nlocor}}
\begin{figure}
	\includegraphics[width=0.4\textwidth]{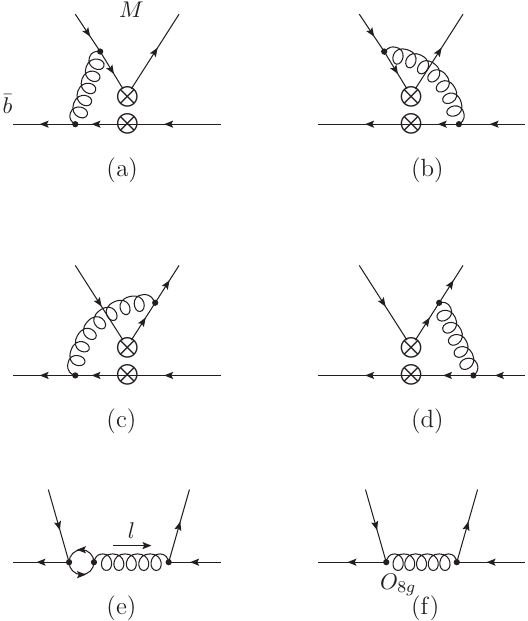}
	\caption{\label{fig:nlodiagrams} Diagrams of NLO QCD corrections: the diagrams (a), (b), (c) and (d) represent the vertex corrections, diagram (e) illustrates the quark-loop corrections, and diagram (f) depicts the chromomagnetic penguin corrections. }
\end{figure}

In addition to the LO calculation, we incorporate several important NLO corrections in this work, concerning the running coupling constant $\alpha_s$, the Wilson coefficients $C_i$, and selected topological diagrams. Specifically, we consider NLO corrections from three factorizable emission diagrams: vertex corrections, quark-loop corrections, and chromomagnetic penguin corrections, as depicted in Fig.~\ref{fig:nlodiagrams}. These diagrams make significant impacts on reducing the scale dependence of the Wilson coefficients, providing potential correction effects to the strong phase, and influencing the $CP$ violation parameters dramatically. The NLO QCD corrections to the nonfactorizable and annihilation diagrams are subdominant and can be neglected at the present precision order.

For convenience, we define the combinations of the Wilson coefficients as
\begin{align} \label{eq:Wilsona}
	a_{1,2}(\mu)=&C_{2,1}(\mu)+\frac{C_{1,2}(\mu)}{N_c}, \nonumber \\
    a_{i,i+1}(\mu)=&C_{i,i+1}(\mu)+\frac{C_{i+1,i}(\mu)}{N_c},
\end{align}
with $i=3,5,7,9$. 
Moreover, we organize the Wilson coefficients as
\begin{align}
	C_{3(4)}^q &= C_{3(4)} + \frac{3}{2}e_q C_{9(10)}, \ a_{3(4)}^q = a_{3(4)} + \frac{3}{2}e_q a_{9(10)},  \nonumber  \\
	C_{5(6)}^q &= C_{5(6)} + \frac{3}{2}e_q C_{7(8)}, \ a_{5(6)}^q = a_{5(6)} + \frac{3}{2}e_q a_{7(8)},
\end{align}
where $e_q$ indicates the electric charge of quark $q$. Since the NLO diagrams shown in Fig.~\ref{fig:nlodiagrams} do not involve the end-point singularity, the amplitudes in the PQCD approach should be the same as those in the QCDF framework \cite{Beneke:2001ev}. 
For the vertex corrections, the gluon should connect any two possible quark lines within the four-quark operators. The effect of these corrections can be represented by the modification of the Wilson coefficients:
\begin{align}
	a_{1,2}(\mu)&\to a_{1,2}(\mu)+\dfrac{\alpha_s(\mu)}{4\pi}C_F\dfrac{C_{1,2}(\mu)}{N_c}V_{1,2}(M), \nonumber\\
    a_{i,i+1}(\mu)&\to a_{i,i+1}(\mu)+\dfrac{\alpha_s(\mu)}{4\pi}C_F\dfrac{C_{i+1,i}(\mu)}{N_c}V_{i,i+1}(M),
\end{align}
with $i=3,5,7,9$. 
Here, the function $V$ depends on the specific type of the final-state meson $M$ that is emitted from the effective vertex as denoted in Fig.~\ref{fig:nlodiagrams}.
When the meson $M$ is a pseudoscalar meson, the function is given by \cite{Beneke:1999br,Beneke:2000ry,Beneke:2001ev,Li:2005kt}
\begin{align}
	V_i(M) =& 12\ln(m_b/\mu) - 18 + \dfrac{2\sqrt{2N_c}}{f_M}\int_0^1dx\phi_M^A(x)g(x), \nonumber \\
		&\hspace{12em} \text{for } i = 1-4,9,10  \nonumber \\
	V_i(M) =& -12\ln(m_b/\mu) + 6 -\dfrac{2\sqrt{2N_c}}{f_M}\int_0^1dx\phi_M^A(x)g(\bar{x}), \nonumber \\
		&\hspace{12em} \text{for } i = 5,7 \nonumber \\
	V_i(M) =& \dfrac{2\sqrt{2N_c}}{f_M}\int_0^1dx\phi_M^P(x)\left[h(x)-6\right], \quad \text{for } i=6,8 \label{eq:vcv}
\end{align}
where $f_M$ is the decay constant, $\phi_M^{A,P}$ are the LCDAs, and the variable $\bar{x}$ is defined by $\bar{x}=1-x$.
The function $V$ associated with the vector meson can be obtained by substituting the LCDAs: $\phi_M^{A,P}\to \phi_V, \phi_V^s$. The hard kernel functions appearing above are defined as
\begin{align}
	g(x)=&3\left(\dfrac{1-2x}{1-x}\ln x-i\pi\right)+\biggl[2\text{Li}_2(x)-\ln^2x \nonumber \\
		&+\dfrac{2\ln x}{1-x}-(3+2i\pi)\ln x-(x\leftrightarrow 1-x)\biggr], \\
	h(x)=&2\text{Li}_2(x)-\ln^2x-(1+2i\pi)\ln x-(x\leftrightarrow 1-x).
\end{align}

Concerning the quark-loop and chromomagnetic penguin corrections, their effective Hamiltonians are given by \cite{Li:2005kt}
\begin{align} \label{eq:heffqlmp}
	H_{\text{eff}}^{\text{ql}} = &-\sum_{q=u,c,t}\sum_{q^\prime}\frac{G_F}{\sqrt{2}}V_{qb}V_{qd}^* \frac{\alpha_s(\mu)}{2\pi}C^{(q)}(\mu,l^2) \nonumber \\
		&\quad\times \left(\bar{d}\gamma_\mu(1-\gamma_5)T^ab\right) \left(\bar{q}^\prime\gamma^\mu T^aq^\prime\right), \nonumber \\
	H_{\text{eff}}^{\text{mp}} = &-\frac{G_F}{\sqrt{2}}V_{tb}V_{td}^* C_{8g}(\mu)O_{8g}(\mu),
\end{align}
where $l^2$ denotes the invariant mass of the gluon connected to the quark loop, as labeled in Fig.~\ref{fig:nlodiagrams}.
In the $b\to s$ transition, the $d$ quark within the $H_{\text{eff}}^{\text{ql}(\text{mp})}$ should be replaced by the $s$ quark.
The function $C^{(q)}$ for the $q$-quark loop is represented by
\begin{align}
	C^{(q)}(\mu,l^2)&=\left[G^{(q)}(\mu,l^2)-2/3\right]C_2(\mu), \nonumber \\
		&\hspace{10em} \text{for } q=u,c \nonumber \\
    C^{(q)}(\mu,l^2)&=-\left[G^{(s)}(\mu,l^2)-2/3\right]C_3(\mu) \nonumber \\
		&-\sum\limits_{q^{\prime\prime}=u,d,s,c}G^{(q^{\prime\prime})}(\mu,l^2)\left[C_4(\mu)+C_6(\mu)\right], \nonumber \\
		&\hspace{10em} \text{for } q=t,
\end{align}
with
\begin{align}
	&G^{(q)}=-4\int_0^1dxx(1-x)\ln\frac{m_q^2-x(1-x)l^2-i\epsilon}{\mu^2}.
\end{align}
Unlike in the case of vertex corrections, the invariant mass of the virtual gluon in the quark-loop diagram simultaneously involves all three mesons participating in the process. Consequently, the expressions for this diagrams must be calculated separately. The results for the quark-loop diagram with $q=t$ are
\begin{align} \label{eq:ql}
	&\mathcal{M}_{\mathrm{ql},P_1V_2}^{(t)} \nonumber \\
	&\textstyle =16\pi^2C_Fm_B^2f_B \left(\frac{C_F}{4\pi N_c}\right)
		\int k_{0\perp}dk_{0\perp} \int dx_0 dx_1 dx_2 \nonumber \\
	&\textstyle \ \times \int b_0db_0b_1db_1 G_B(x_0,b_0,k_{0\perp})
		\{C^{(t)}(t_q,l_{q}^2) \alpha_s^2(t_q) \nonumber \\
	&\textstyle \ \times [((2-x_1)E_q-x_1k_0^z)\phi_{P_1}^A(x_1,b_1)\phi_{V_2}(x_2,0) \nonumber \\
	&\textstyle \  -r_{P_1}((1-2x_1)E_q-k_0^z)(\phi_{P_1}^P(x_1,b_1)-\phi_{P_1}^T(x_1,b_1)) \nonumber \\
		&\textstyle \  \times \phi_{V_2}(x_2,0)
	-2r_{V_2}(E_q-k_0^z)\phi_{P_1}^A(x_1,b_1)\phi_{V_2}^s(x_2,0) \nonumber \\
	&\textstyle \  -2r_{P_1} r_{V_2} ((3-x_1)E_q+(1-x_1)k_0^z) \phi_{P_1}^P(x_1,b_1) \nonumber \\
		&\textstyle \  \times \phi_{V_2}^s(x_2,0)
		-2r_{P_1} r_{V_2}((1-x_1)E_q-(1+x_1)k_0^z) \nonumber \\ &\textstyle \  \times \phi_{P_1}^T(x_1,b_1)\phi_{V_2}^s(x_2,0)]
		h_e(x_0,1-x_1,b_0,b_1) S_t(x_1) \nonumber \\
	&\textstyle \  \times \exp[-S_B(t_q)-S_{P_1}(t_q)]
		+C^{(t)}(t_q^\prime,l_{q}^{\prime 2}) \alpha_s^2(t_q^\prime)
		2r_{P_1} \nonumber \\ & \textstyle \  \times (E_q+k_0^z)\phi_{P_1}^P(x_1,b_1) (\phi_{V_2}(x_2,0) - 2 r_{V_2} \phi_{V_2}^s(x_2,0)) \nonumber \\
	&\textstyle \ \times h_e(1-x_1,x_0,b_1,b_0) S_t(x_0)
		\exp[-S_B(t_q^\prime)-S_{P_1}(t_q^\prime)]\},
\end{align}
where the relevant constants, functions, and limits of integration are identical to those in Eq.~(\ref{eq:fe}).
Analogous to the pseudoscalar meson, the LCDAs and chiral parameter for the vector meson are represented by $\phi_{V_2}^{\_,s,t}$ and $r_{V_2}=m_{V_2}/m_B$.
The expressions for diagrams involving the $u$- and $c$-quark loop can be derived by direct substitution of quark flavor.
For the sake of conciseness, the result $\mathcal{M}_{\mathrm{mp}}$ for the chromomagnetic penguin diagram, along with the details of the scale $t$ and gluon's invariant mass $l^2$, are included in Appendix~\ref{app:fullamp}.

\subsection{Contributions of soft form factors and color-octet quark-antiquark components \label{nonpertcor}}
\begin{table*}
  \caption{\label{tab:fevsalphas}
  Numerical results for the amplitude (in units of $10^{-8}$) of $B^+ \to \pi^0 \rho^+$ contributed by the factorizable emission diagrams Fig.~\ref{fig:eightdiagram} (a) and (b) as a function of the running coupling.  The CKM and Wilson coefficients for the amplitude $F_{e,\pi\rho}$ of this channel is $V_{ub}^*V_{ud}(C_1/3+C_2)-V_{tb}^*V_{td}(C_3/3+C_4+C_9/3+C_{10})$. The phases in these values originate from the CKM matrix elements. 
  }
  \renewcommand\arraystretch{1.5}
  \footnotesize
  \begin{ruledtabular}
    \begin{tabular}{cccccccccccc}
      $\alpha_s/\pi$	& $\le 0.1$ & $\le 0.2$ & $\le 0.3$ & $\le 0.4$ & $\le 0.5$ & $\le 0.6$ & Total \\
      \hline
      $F_{e,\pi\rho}$ 	& $1.21\text{Exp}(0.83\pi i)$ & $3.60\text{Exp}(0.82\pi i)$ & $4.59\text{Exp}(0.82\pi i)$ & $5.14\text{Exp}(0.81\pi i)$ & $5.52\text{Exp}(0.81\pi i)$ & $5.80\text{Exp}(0.81\pi i)$ & $7.32\text{Exp}(0.72\pi i)$\\
      $V_u$ part  		& $1.20\text{Exp}(0.86\pi i)$ & $3.55\text{Exp}(0.86\pi i)$ & $4.51\text{Exp}(0.86\pi i)$ & $5.05\text{Exp}(0.86\pi i)$ & $5.41\text{Exp}(0.86\pi i)$ & $5.67\text{Exp}(0.86\pi i)$ & $6.44\text{Exp}(0.86\pi i)$\\
      $V_t$ part  		& $0.11\text{Exp}(0.37\pi i)$ & $0.46\text{Exp}(0.37\pi i)$ & $0.67\text{Exp}(0.37\pi i)$ & $0.82\text{Exp}(0.37\pi i)$ & $0.94\text{Exp}(0.37\pi i)$ & $1.03\text{Exp}(0.37\pi i)$ & $3.26\text{Exp}(0.37\pi i)$\\
	  $|F_{e,\pi\rho}|$ in \% 	& $16.5\%$ & $49.2\%$ & $62.7\%$ & $70.2\%$ & $75.4\%$ & $79.2\%$ & $100.0\%$ \\
	  $V_u$ part in \%			& $18.6\%$ & $55.2\%$ & $70.1\%$ & $78.5\%$ & $84.0\%$ & $88.1\%$ & $100.0\%$ \\
	  $V_t$ part in \%			& $3.4 \%$ & $14.1\%$ & $20.6\%$ & $25.2\%$ & $28.8\%$ & $31.6\%$ & $100.0\%$
    \end{tabular}
  \end{ruledtabular}
\end{table*}

Based on the method presented in Secs. \ref{subsec:loamp} and \ref{subsec:nlocor}, the numerical amplitudes can be obtained for various decay channels. With these calculations, we are capable of checking the reliability of the perturbative calculation. As an example, we present in Table~\ref{tab:fevsalphas} the contributions from different ranges of $\alpha_s/\pi$ value to the amplitudes of the factorizable emission diagrams for $B^+\to \pi^0\rho^+$ decay, which correspond to Figs.~\ref{fig:eightdiagram} (a) and (b).
The numerical results show that more than 50\% of the contributions to the amplitude of Figs.~\ref{fig:eightdiagram} (a) and (b) originate from the soft region with $\alpha_s(\mu)/\pi>0.2$, where $\alpha_s(\mu)/\pi\sim 0.2$ can be viewed as the boundary between the soft and hard contributions. In the soft region, where the strong coupling constant is large, perturbative calculation is not applicable. This suggests the inadequacy of applying the perturbation theory in a naive manner. Actually, the significant contribution from the soft region can be attributed to the end-point behavior of the $B$ meson wave function employed in this study. Compared with the conventional wave function model, the wave function derived from the relativistic potential model exhibits a slower decrease as the momentum fractions of the quark and antiquark within the $B$ meson tend to zero.

To enhance the reliability of the PQCD calculation, we can introduce a critical infrared cutoff scale $\mu_c$ to separate the hard and soft contributions. The contribution with the scale above the scale $\mu_c$ is computed using perturbation theory, as outlined in the preceding Secs. A and B, whereas the contribution with the scale below $\mu_c$ is treated as soft and must be determined through nonperturbative methods. In this work, we estimate the nonperturbative contributions with the help of the results from QCD LCSR or by constraining them with available experimental data.

We performed the numerical analysis for the contributions of all the other diagrams, and found that the nonperturbative contribution of the factorizable diagrams -- specifically diagrams (a), (b), (g), and (h) in Fig.~\ref{fig:eightdiagram} -- is not so small, while for the nonfactorizable diagrams, the soft contribution is very small, which could be safely neglected. Therefore, we need to introduce soft form factors for factorizable emission and annihilation diagrams to encompass these soft contributions. In the $B\to P$ and $B\to V$ transitions, the transition form factors $F_{+,0}$, $A_{0,1,2,3}$, and $V$ are defined by
\begin{align} \label{eq:BMff}
	\langle P|&\bar{q}\gamma_\mu(1-\gamma_5)b|\bar{B}\rangle \nonumber \\
		=& \left(p_B+p_P-\dfrac{m_B^2-m_P^2}{q^2}q\right)_\mu F_+(q^2) \nonumber \\
		&+ \dfrac{m_B^2-m_P^2}{q^2}q_\mu F_0(q^2), \\
	\langle V|&\bar{q}\gamma_\mu(1-\gamma_5)b|\bar{B}\rangle \nonumber \\
		=&\dfrac{2\epsilon_{\mu\nu\rho\sigma}}{m_B+m_V}\epsilon_V^\nu p_B^\rho p_V^\sigma V(q^2) +i\dfrac{\epsilon_V\cdot q}{q^2}2m_Vq_\mu A_0(q^2) \nonumber \\
		&+i\left[\epsilon_{V\mu}(m_B+m_V)A_1(q^2) - \dfrac{\epsilon_V\cdot q}{q^2}2m_Vq_\mu A_3(q^2) \right. \nonumber \\
		&-\left.\dfrac{\epsilon_V\cdot q}{m_B+m_V}\left(p_B+p_V\right)_\mu A_2(q^2)\right],
\end{align}
where $q$ is the momentum transfer $q = P_B-P_{P(V)}$, and $\epsilon_V$ is the polarization vector. The relevant form factors for $B\to PV$ decays are $F_+$ and $A_0$. We can define the hard and soft form factors $h$ and $\xi$ for the transitions $B\to P$ and $B\to V$ as
\begin{align} \label{eq:BMff2}
	F_+^{BP} &= h_+^{BP} + \xi_{+}^{BP},\quad A_0^{BV} = h_{A0}^{BV} + \xi_{A0}^{BV}.
\end{align}
The hard contributions $h_+^{BP}$ and $h_{A0}^{BV}$ are associated with the factorizable emission diagrams and can be perturbatively calculated with the energy scale larger than $\mu_c$. The total transition form factors $F_+^{BP}$ and $A_0^{BV}$, derived from LCSR, are adopted as input parameters. Subsequently, with these LCSR inputs, we can evaluate the soft form factors $\xi_{+}^{BP}$ and $\xi_{A0}^{BV}$.

Similarly, we define the production form factor corresponding to the factorizable annihilation diagrams as
\begin{align} \label{eq:M1M2ff}
	\langle M_1M_2 |(S+P)|0 \rangle = -\dfrac{1}{2}\sqrt{\mu_1\mu_2}F^{M_1M_2},
\end{align}
where $M_{1,2}$ denote the final-state mesons and $\mu_{1,2}$ symbolize their chiral masses. The chiral masses for pseudoscalar and vector mesons differ from the chiral parameters $r_{P(V)}$ solely by a multiplicative factor, such that $\mu_{P(V)} = m_B r_{P(V)}$. In contrast to the transition form factors, there are no reliable theoretical predictions for these production form factors introduced here up to now. Therefore, they are treated as nonperturbative input parameters in this work, and their values are determined by the global fitting to experimental data.

Furthermore, in our earlier works trying to resolve the $B\to\pi\pi$ and $\pi K$ puzzles, we found that the contributions from the quark-antiquark pair in the color-octet state are crucial \cite{Wang:2022ihx,Lu:2022hbp}. The color-octet contribution also plays an important role in explaining the branching ratios and $CP$ violations for numerous charmless $B\to PP$ decays \cite{Lu:2024jzn}. In this work on $B\to PV$ decays, this color-octet contribution should also be considered.

The concept of the color-octet contribution can be explained as follows, although the associated calculations are quite challenging. One generally expects color-singlet mesons in the final state. However, the QCD interaction does not exclude the formation of color-octet quark-antiquark pairs at short distances. As long as these color-octet pairs convert into their corresponding mesons by exchanging soft gluons at long distances, this process should indeed contribute to the decay amplitude. To estimate the magnitude of this contribution, we divide the process into two distinct stages. In the short-distance region, the transition from the initial meson to two color-octet quark-antiquark pairs is under control by a perturbative treatment. The Feynman diagrams corresponding to the color-octet pairs are analogous to those of their color-singlet counterparts, i.e., the diagrams in Fig.~\ref{fig:eightdiagram}, except that the color-singlet quark-antiquark pairs in the final state are replaced by color-nonsinglet states. The only difference resides in the color factor associated with each diagram in Fig.~\ref{fig:eightdiagram}. With the relation for color SU(3) generators $T^a_{ik}T^a_{jl}=-\delta_{ik}\delta_{jl}/{2N_c}+\delta_{il}\delta_{jk}/2$, we can derive the color factor. By assuming the LCDA of a quark-antiquark pair in the color-octet state is the same as that for the color-singlet meson state, one can calculate the amplitudes for various topological diagrams involving color-octet quark-antiquark pairs in the same way as is done for color-singlet states. The details of the method for analyzing the color factor can be found in our previous works in Refs. \cite{Wang:2022ihx,Lu:2024jzn}. By substituting the color factors, these short-distance amplitudes can be expressed as follows
\begin{align} \label{eq:fm8}
	F_{e,M_1M_2}^{(P,R),8} &\equiv 2N_c^2 F_{e,M_1M_2}^{(P,R),a} - \dfrac{N_c}{C_F} F_{e,M_1M_2}^{(P,R),b}, \nonumber\\
    \mathcal{M}_{e,M_1M_2}^{(P,R),8} &\equiv 2 N_c^2 \mathcal{M}_{e,M_1M_2}^{(P,R),c} - \dfrac{N_c}{C_F} \mathcal{M}_{e,M_1M_2}^{(P,R),d}, \nonumber\\
    \mathcal{M}_{e,M_1M_2}^{(P,R),8^\prime} &\equiv \dfrac{N_c^2}{C_F} \mathcal{M}_{e,M_1M_2}^{(P,R)}, \
	\mathcal{M}_{a,M_1M_2}^{(P,R),8} \equiv -\dfrac{N_c}{C_F} \mathcal{M}_{a,M_1M_2}^{(P,R)},\nonumber\\
	F_{a,M_1M_2}^{(P,R),8} &\equiv -\dfrac{N_c^2}{C_F} F_{a,M_1M_2}^{(P,R)},
\end{align}
where the definitions of $F_e$, $M_e$, $M_a$ and $F_a$ are the same as those given previously before Eq.~(\ref{eq:fe}). To distinguish them from the results for color-singlet states, we use the superscripts $8$ and $8^\prime$ to label the amplitudes of color-octet states, and the superscripts $a$, $b$, $c$, and $d$ on the right-hand side of the above equation indicate the amplitudes for the relevant diagrams (a), (b), (c) and (d) shown in Fig.~\ref{fig:eightdiagram}. $F_{e,M_1M_2}^{(P,R)}$, $\mathcal{M}_{e,M_1M_2}^{(P,R)}$, $\mathcal{M}_{a,M_1M_2}^{(P,R)}$ and $F_{a,M_1M_2}^{(P,R)}$ are the color-singlet amplitudes relevant to the diagrams in Fig.~\ref{fig:eightdiagram}, which are given in Appendix~\ref{app:fullamp}, and the meaning of the superscripts $P$ and $R$ can be also found in Appendix~\ref{app:fullamp}.

At long distance, color-octet quark-antiquark pairs convert into color-singlet mesons through the exchange of soft gluons. Since the energy scale is within the soft region of QCD,  perturbation calculation breaks down. Given the nonperturbative property involved in this transformation, we introduce the associated parameters $Y^8_{F,M}$ to characterize its effect in factorizable and nonfactorizable diagrams. Their values must be derived by fitting experimental results.



By combining the contributions from each component with the corresponding Wilson coefficients, we can obtain the expression of the decay amplitude for the color-octet contribution in each decay channel. We take $B^-\to \rho^-\pi^0$ decay as an example, and its decay amplitude can be represented as follows
\begin{align} \label{eq:amprhopi}
	&\sqrt{2}\mathcal{M}\left(B^-\to \rho^-\pi^0\right) \nonumber \\
	&= V_{ub}V_{ud}^*
		\Big[a_2F_{e,\rho\pi} + C_2\mathcal{M}_{e,\rho\pi}/N_c
		+ a_1F_{e,\pi\rho} \nonumber \\ &\ + C_1 \mathcal{M}_{e,\pi\rho}/N_c
		+ a_1 F_{a,\rho\pi} + C_1 \mathcal{M}_{a,\rho\pi}/N_c
		- a_1 F_{a,\pi\rho} \nonumber \\ &\ - C_1 \mathcal{M}_{a,\pi\rho}/N_c
		+ a_{2,\text{vc}}F_{e,\rho\pi} -\mathcal{M}_{\text{ql},\rho\pi}^{(u)}
		+ a_{1,\text{vc}} F_{e,\pi\rho} \nonumber \\ &\ + \mathcal{M}_{\text{ql},\pi\rho}^{(u)}
		- 2 \left(a_2^{\mu_c} + a_{2,\text{vc}}^{\mu_c}\right)f_\pi\xi^{B\rho} \nonumber \\
		&\ - 2 \left(a_1^{\mu_c} + a_{1,\text{vc}}^{\mu_c}\right) f_\rho^\parallel \xi^{B\pi}
		+ \big(C_2 F_{e,\rho\pi}^8 + C_1F_{e,\pi\rho}^8 \nonumber \\
		&\ + a_1F_{a,\rho\pi}^8 - a_1F_{a,\pi\rho}^8 \big)Y_{F,\rho\pi}^8
		+ \Big(C_2 \mathcal{M}_{e,\rho\pi}^8 + C_1 \mathcal{M}_{e,\rho\pi}^{8^\prime} \nonumber \\
		&\ + C_1\mathcal{M}_{e,\pi\rho}^8 + C_2\mathcal{M}_{e,\pi\rho}^{8^\prime} + C_1\mathcal{M}_{a,\rho\pi}^8 - C_1\mathcal{M}_{a,\pi\rho}^8 \Big) \nonumber \\ &\ \times Y_{M,\rho\pi}^8 \Big]
	+ V_{cb}V_{cd}^*\Big[-\mathcal{M}_{\text{ql},\rho\pi}^{(c)} + \mathcal{M}_{\text{ql},\pi\rho}^{(c)}\Big] \nonumber \\
	&\ - V_{tb}V_{td}^*
		\Big\{\left(a_3^u-a_3^d-a_4^d\right) F_{e,\rho\pi} + \left(a_5^u-a_5^d\right)F_{e,\rho\pi}^R \nonumber \\ &\ - a_6^dF_{e,\rho\pi}^P
		+ \left(C_4^u-C_3^d-C_4^d\right) \mathcal{M}_{e,\rho\pi}/N_c \nonumber \\ &\ - C_5^d \mathcal{M}_{e,\rho\pi}^R/N_c + \left(C_6^u-C_6^d\right)\mathcal{M}_{e,\rho\pi}^P/N_c \nonumber \\
		&\ + a_4^u F_{e,\pi\rho} + a_6^u F_{e,\pi\rho}^P
		+ C_3^u \mathcal{M}_{e,\pi\rho}/N_c + C_5^u \mathcal{M}_{e,\pi\rho}^R/N_c \nonumber \\
		&\ + a_4^u F_{a,\rho\pi} + a_6^u F_{a,\rho\pi}^P + C_3^u \mathcal{M}_{a,\rho\pi}/N_c + C_5^u \mathcal{M}_{a,\rho\pi}^R/N_c \nonumber \\
		&\ - a_4^u F_{a,\pi\rho} - a_6^u F_{a,\pi\rho}^P - C_3^u \mathcal{M}_{a,\pi\rho}/N_c - C_5^u \mathcal{M}_{a,\pi\rho}^R/N_c \nonumber \\
		&\ + \left(a_{3,\text{vc}}^u - a_{3,\text{vc}}^d - a_{4,\text{vc}}^d\right) F_{e,\rho\pi}
		+ \left(a_{5,\text{vc}}^u - a_{5,\text{vc}}^d\right) F_{e,\rho\pi}^R \nonumber \\
		&\ - a_{6,\text{vc}}^d F_{e,\rho\pi}^P - \mathcal{M}_{\text{ql},\rho\pi}^{(t)} - \mathcal{M}_{\text{mp},\rho\pi} \nonumber \\
		&\ + a_{4,\text{vc}}^u F_{e,\pi\rho} + a_{6,\text{vc}}^u F_{e,\pi\rho}^P + \mathcal{M}_{\text{ql},\pi\rho}^{(t)} + \mathcal{M}_{\text{mp},\pi\rho} \nonumber \\
		&\ - 2 \Big[\left(a_3^{u,\mu_c}-a_3^{d,\mu_c}-a_4^{d,\mu_c}\right)
		- \left(a_5^{u,\mu_c} - a_5^{d,\mu_c}\right) \nonumber \\
		&\ + \left(a_{3,\text{vc}}^{u,\mu_c} - a_{3,\text{vc}}^{d,\mu_c} - a_{4,\text{vc}}^{d,\mu_c} \right)
		- \left(a_{5,\text{vc}}^{u,\mu_c} - a_{5,\text{vc}}^{d,\mu_c}\right) \nonumber \\
		&\ - 2r_\pi \left(-a_6^{d,\mu_c} - a_{6,\text{vc}}^{d,\mu_c}\right)\Big] f_\pi\xi^{B\rho} \nonumber \\
		&\ -2 \Big[\left(a_4^{u,\mu_c} + a_{4,\text{vc}}^{u,\mu_c}\right)
		- 2r_\rho \left(a_6^{u,\mu_c} + a_{6,\text{vc}}^{u,\mu_c}\right) \Big]f_\rho^\parallel \xi^{B\pi} \nonumber \\
		&\ -2\chi_B f_B \left(a_6^{u,\mu_c} + a_6^{u,\mu_c} \right) \sqrt{r_\rho r_\pi} \xi^{\rho\pi} \nonumber \\
		&\ + \Big[\left(C_4^u-C_3^d-C_4^d\right)F_{e,\rho\pi}^8 + \left(C_6^u-C_6^d\right)F_{e,\rho\pi}^{R,8} \nonumber \\ &\ - C_5^dF_{e,\rho\pi}^{P,8}
		+ C_3^uF_{e,\pi\rho}^8 + C_5^uF_{e,\pi\rho}^{P,8} \nonumber \\
		&\ + a_4^uF_{a,\rho\pi}^8 + a_6^uF_{a,\rho\pi}^{P,8}
		- a_4^uF_{a,\pi\rho}^8 - a_6^uF_{a,\pi\rho}^{P,8} \Big] Y_{F,\rho\pi}^8 \nonumber \\
		&\ + \Big[\left(C_4^u-C_3^d-C_4^d\right)\mathcal{M}_{e,\rho\pi}^8 - C_5^d\mathcal{M}_{e,\rho\pi}^{R,8} \nonumber \\ &\ + \left(C_6^u-C_6^d\right) \mathcal{M}_{e,\rho\pi}^{P,8}
		+ \left(C_3^u-C_4^d-C_3^d\right)\mathcal{M}_{e,\rho\pi}^{8^\prime} \nonumber \\ &\ - C_6^d\mathcal{M}_{e,\rho\pi}^{R,8^\prime} + \left(C_5^u-C_5^d\right)\mathcal{M}_{e,\rho\pi}^{P, 8^\prime}
		+ C_3^u\mathcal{M}_{e,\pi\rho}^8 \nonumber \\ & \ + C_5^u\mathcal{M}_{e,\pi\rho}^{R,8} + C_4^u\mathcal{M}_{e,\pi\rho}^{8^\prime} + C_6^u\mathcal{M}_{e,\pi\rho}^{R,8^\prime}
		+ C_3^u\mathcal{M}_{a,\rho\pi}^8 \nonumber \\ &\ + C_5^u\mathcal{M}_{a,\rho\pi}^{R,8}
		- C_3^u\mathcal{M}_{a,\pi\rho}^8 - C_5^u\mathcal{M}_{a,\pi\rho}^{R,8} \Big] Y_{M,\rho\pi}^8\Big\},
\end{align}
with the factor $\chi_B = 1.388$ originating from the integration of $B$ meson wave function, which can be found in Eq. (\ref{chiB}).

\subsection{SU(3) symmetry and symmetry-breaking analysis for the soft parameters \label{app:su3}}

We have introduced the parameters for the production form factors and color-octet matrix elements, whose values require global fitting from experimental data. To avoid compromising the predictive power of the theoretical framework with an excess of parameters, we invoke the flavor SU(3) symmetry. The symmetry interrelates the various decay channels, thereby allowing the parameters introduced above for each individual channel to be expressed through a more limited set of SU(3) parameters.

Within the framework of flavor SU(3) symmetry, the light pseudoscalar and vector mesons can be organized into nonets as \cite{Xu:2013dta,He:2018joe,Wang:2022yyn}
\begin{equation}
	P = \left(
		\begin{array}{ccc}
			\frac{\pi^0+\eta_q}{\sqrt{2}}	& \pi^+ 							& K^+ 	\\
			\pi^- 							& \frac{-\pi^0+\eta_q}{\sqrt{2}} 	& K^0 	\\
			K^-								& \bar{K}^0							& \eta_s
		\end{array}
	\right),
\end{equation}
and
\begin{equation}
	V = \left(
		\begin{array}{ccc}
			\frac{\rho^0+\omega}{\sqrt{2}}	& \rho^+							& K^{*+} 	\\
			\rho^-							& \frac{-\rho^0+\omega}{\sqrt{2}}	& K^{*0}	\\
			K^{*-}							& \bar{K}^{*0}						& \phi		
		\end{array}
	\right),
\end{equation}
respectively.

According to SU(3) symmetry, within each nonet, the particles are symmetric under the strong interaction. However, for mesons in different nonets $P$ and $V$, the symmetry is poor, which can be seen from the large difference between the masses of pseudoscalar and vector mesons. The vector mesons are different from the pseudoscalars by a spin flip, which can cause large symmetry breaking. Therefore, we only consider SU(3) symmetry with final mesons within their own nonets in $B\to PV$ decays. Nevertheless, the relation between the flavor symmetries of $B\to PV$ and $B\to PP$ is still interesting, which shall be considered elsewhere in the future.

In light of the mass of the $s$ quark being much heavier than that of the $u$ and $d$ quarks, which leads to the breaking of flavor SU(3) symmetry, a matrix $W$ can be introduced to describe the symmetry-breaking effect, defined as \cite{Xu:2013dta,He:2018joe,Wang:2022yyn}
\begin{equation}
	W = \left(
		\begin{array}{ccc}
			0 & 0 & 0 \\
			0 & 0 & 0 \\
			0 & 0 & 1
		\end{array}
	\right).
\end{equation}

The color-octet parameters describe the transition of color-octet quark-antiquark pairs into color-singlet mesons, which is a process mediated by exchanging soft gluons at long distance. For simplicity, we assume that the soft gluons will not change the spin of the quark-antiquark pairs during the transition process. Consequently, a pseudoscalar or vector quark-antiquark pair in the color-octet state remains to be a pseudoscalar or vector state after the color transformation. Within the framework of flavor SU(3) symmetry, the effective Hamiltonian for the $P_8V_8 \to P_1V_1$ transition can be formulated as
\begin{equation} \label{eq:coheff0}
	H_0^{a(b)}= C_0^{a(b)} (P_8)^i_j (P_1^{\dagger})^j_i \cdot (V_8)^k_l (V_1^{\dagger})^l_k,
\end{equation}
where the subscripts $8$ and $1$, respectively, denote the color-octet and -singlet states, while the superscripts $a$ and $b$ serve to distinguish the contributions associated with the parameters $Y_F^8$ and $Y_M^8$.
The coefficient $C_0^{a(b)}$ represents the effective coupling constant.
To account for the symmetry-breaking effects, it is necessary to consider all possible modes of breaking at a given order.
At the first order, two distinct modes of breaking emerge, whose corresponding Hamiltonian terms are given by
\begin{eqnarray} \label{eq:coheff1}
	H_1^{1,a(b)} &=& C_1^{1,a(b)} \Big\{\big[W_j^i (P_8)_m^j (P_1^{\dagger})_i^m \big]\big[(V_8)_l^k (V_1^{\dagger})_k^l\big] \nonumber \\
		&& \quad + \big[(P_8)_s^r W_t^s (P_1^{\dagger})_r^t\big] \big[(V_8)_l^k (V_1^{\dagger})_k^l\big] \Big\}, \nonumber \\
	H_1^{2,a(b)} &=& C_1^{2,a(b)} \Big\{\big[(P_8)_j^i (P_1^{\dagger})_i^j\big] \big[W_l^k (V_8)_m^l (V_1^{\dagger})_k^m \big] \nonumber \\
		&& \quad + \big[(P_8)_j^i (P_1^{\dagger})_i^j\big] \big[(V_8)_s^r W_t^s (V_1^{\dagger})_r^t\big] \Big\}.
\end{eqnarray}
At the second order, the Hamiltonian terms that include all possible symmetry-breaking modes are listed as follows:
\begin{eqnarray} \label{eq:coheff2}
	H_2^{1,a(b)} &=& C_2^{1,a(b)} \Big\{\big[W_j^i (P_8)_m^j (P_1^{\dagger})_i^m\big] \big[W_l^k (V_8)_n^l (V_1^{\dagger})_k^n\big] \nonumber \\
		&&\quad + \big[(P_8)_s^r W_t^s (P_1^{\dagger})_r^t\big] \big[(V_8)_v^u W_w^v (V_1^{\dagger})_u^w\big] \Big\}, \nonumber \\
	H_2^{2,a(b)} &=& C_2^{2,a(b)} \Big\{\big[W_j^i (P_8)_m^j (P_1^{\dagger})_i^m\big] \big[(V_8)_l^k W_n^l (V_1^{\dagger})_k^n\big] \nonumber \\
		&&\quad + \big[(P_8)_s^r W_t^s (P_1^{\dagger})_r^t\big] \big[W_v^u (V_8)_w^v (V_1^{\dagger})_u^w\big] \Big\}, \nonumber \\
	H_2^{3,a(b)} &=& C_2^{3,a(b)} \big[W_j^i (P_8)_m^j W_n^m (P_1^{\dagger})_i^n \big] \big[(V_8)_l^k (V_1^{\dagger})_k^l\big], \nonumber \\
	H_2^{4,a(b)} &=& C_2^{4,a(b)} \big[(P_8)_j^i (P_1^{\dagger})_i^j\big] \big[W_l^k (V_8)_m^l W_n^m (V_1^{\dagger})_k^n \big]. \nonumber \\
\end{eqnarray}
In Eqs.~(\ref{eq:coheff0})-(\ref{eq:coheff2}), the numerical subscripts, $0,\ 1,\ 2$, labeled on the Hamiltonian terms and effective couplings indicate the order of the symmetry breaking, whereas the numerical superscripts distinguish the contributions from different terms at the same order.
It is worth noting that the strong interaction conserves $CP$ parity.
In each of the aforementioned Hamiltonian terms, the $CP$ symmetry has been explicitly preserved.

In the $B\to PV$ decay, the production form factor is defined by the matrix element induced by the pseudoscalar current.
For the cases of flavor SU(3) symmetry,
the corresponding terms of the effective Hamiltonian are given
by
\begin{equation} \label{eq:pffheff0}
	H_0^c = C_0^c (P_j^i V_k^j T_i^k + V_m^l P_n^m T_l^n),
\end{equation}
which is the leading-order Hamiltonian that keeps SU(3) symmetry. The first-order symmetry-breaking terms are
\begin{eqnarray} \label{eq:pffheff1}
	H_1^{1,c} &=& C_1^{1,c} (W_j^i P_k^j V_l^k T_i^l + V_n^m P_s^n W_t^s T_m^t), \nonumber \\
	H_1^{2,c} &=& C_1^{2,c} (P_j^i W_k^j V_l^k T_i^l + V_n^m W_s^n P_t^s T_m^t), \nonumber \\
	H_1^{3,c} &=& C_1^{3,c} (P_j^i V_k^j W_l^k T_i^l + W_n^m V_s^n P_t^s T_m^t),
\end{eqnarray}
and the second-order symmetry-breaking terms are
\begin{eqnarray} \label{eq:pffheff2}
	H_2^{1,c} &=& C_2^{1,c} (W_j^i P_k^j W_l^k V_m^l T_i^m + V_s^r W_t^s P_u^t W_v^u T_r^v), \nonumber \\
	H_2^{2,c} &=& C_2^{2,c} (W_j^i P_k^j V_l^k W_m^l T_i^m + W_s^r V_t^s P_u^t W_v^u T_r^v), \nonumber \\
	H_2^{3,c} &=& C_2^{3,c} (P_j^i W_k^j V_l^k W_m^l T_i^m + W_s^r V_t^s W_u^t P_v^u T_r^v), \nonumber \\
\end{eqnarray}
where $T_j^i = \bar{q}_j \gamma_5 q_i^\prime $ denotes the pseudoscalar current with $q_{1,2,3}^{(\prime)}=u, d, s$ quark.
In Eqs.~(\ref{eq:pffheff0})-(\ref{eq:pffheff2}), the superscript $c$ represents the contribution related to the production form factor.

Based on the effective Hamiltonians shown above, we can derive the expressions of the color-octet parameters and production form factors among the various $B\to PV$ decay channels in terms of the relevant effective couplings $C_{0,1,2}$. To demonstrate the magnitude of symmetry breaking, the ratios $R_{1,2}$ with respect to $C_0$ are employed, rather than direct expressions in terms of the $C_{1,2}$. The consequent parametrization of the nonperturbative parameters for various decay channels is tabulated in Table.~\ref{tab:su3relation}.


\begin{table*}
  \caption{\label{tab:su3relation}
	Expressions for production form factors and color-octet parameters are given in terms of the flavor SU(3) symmetry and symmetry-breaking parameters $C$ and $R$, where the subscripts $0,1,2$ indicate the order of SU(3) symmetry breaking, and the superscripts $a$, $b$, $c$ are for parameters $Y_{8F}$,  $Y_{8M}$ and the production form factor $F_{M_1M_2}$, respectively.
  }
  \renewcommand\arraystretch{1.5}
  \begin{ruledtabular}
    \begin{tabular}{cccccccccccc}
      Channel & $\sqrt{\mu_{1}\mu_{2}}F_{M_1M_2}$ & $Y^8_{F(M),M_1M_2}$ \\
      \hline
      $\pi\rho,\ \eta_q\rho,\ \pi\omega$ & $C_0^{c}$											 & $C_0^{a(b)}$                                                                                  \\
      $\pi K^*,\ \eta_q K^*$             & $C_0^{c}\left(1+R_1^{3,c}\right)$					 & $C_0^{a(b)}\left(1+R_1^{2,a(b)}\right)$                                                       \\
      $K \rho,\ K\omega$                 & $C_0^{c}\left(1+R_1^{1,c}\right)$					 & $C_0^{a(b)}\left(1+R_1^{1,a(b)}\right)$                                                       \\
      $K K^*$                            & $C_0^{c}\left(1+R_1^{2,c}\right)$					 & $C_0^{a(b)}\left(1+R_1^{1,a(b)}+R_1^{2,a(b)}+R_2^{2,a(b)}\right)$                             \\
      $\eta_s\rho$                       & 0													 & $C_0^{a(b)}\left(1+2R_1^{1,a(b)}+R_2^{3,a(b)}\right)$                                         \\
      $\pi\phi$                          & 0													 & $C_0^{a(b)}\left(1+2R_1^{2,a(b)}+R_2^{4,a(b)}\right)$                                         \\
      $\eta_s K^*$                       & $C_0^{c}\left(1+R_1^{1,c}+R_1^{2,c}+R_2^{1,c}\right)$ & $C_0^{a(b)}\left(1+2R_1^{1,a(b)}+R_1^{2,a(b)}+R_2^{1,a(b)}+R_2^{2,a(b)}+R_2^{3,a(b)}\right)$  \\
      $K\phi$                            & $C_0^{c}\left(1+R_1^{2,c}+R_1^{3,c}+R_2^{3,c}\right)$ & $C_0^{a(b)}\left(1+R_1^{1,a(b)}+2R_1^{2,a(b)}+R_2^{1,a(b)}+R_2^{2,a(b)}+R_2^{4,a(b)}\right)$  \\
    \end{tabular}
  \end{ruledtabular}
\end{table*}

Due to the relevant CKM matrix elements, the amplitude is divided into three parts, each multiplied by $V_{ub}V_{ud}^*$, $V_{cb}V_{cd}^*$, and $V_{tb}V_{td}^*$. In each part, we consider the contributions arising from LO diagrams, NLO diagrams, soft form factors, and color-octet components. The combinations of Wilson coefficients corresponding to LO and NLO diagrams for $B\to PV$ decay channels are systematically organized in Tables~\ref{tab:wilson1} and \ref{tab:wilson2} in Appendix~\ref{app:wilson}. For color-octet contributions, the relevant combinations of the Wilson coefficients can be deduced from those of the associated LO diagrams. The derivation is also described in Appendix~\ref{app:wilson}. For the contribution of soft transition and production form factors, the Wilson coefficients can be obtained from the relevant factorizable emission and annihilation diagrams. The details are elaborated in our previous works \cite{Lu:2024jzn}. It is necessary to note that the Wilson coefficients associated with the contributions of soft form factors should be taken at the scale $\mu=\mu_c$, because these contributions are from low scale.

\section{Numerical Results and Discussion \label{sec:num}}
Based on the theoretical framework outlined in the preceding sections, we present the specific numerical calculations and discussion in this section. The essential input parameters for this calculation are the masses, decay constants, and chiral parameters for the related mesons. The values of these parameters are shown in Table~\ref{tab:input}.
\begin{table}
  \caption{\label{tab:input} Masses, decay constants, and chiral parameters for $B$ meson, light pseudoscalar and vector mesons. }
  \begin{ruledtabular}
    \begin{tabular}{cccccccccccc}
      Meson & $m (\text{GeV})$ \cite{ParticleDataGroup:2024cfk}
	  		& $f_{P(V)}^{(\parallel/\perp)}(\text{GeV})$ \cite{Ball:2006wn, Ball:2007rt, Sun:2019xyw} & $r_{P,V}$ \\
      \hline
      $B$ 				& $5.279$ 	& $0.210$ 		& $\cdots$	\\
      $\pi$ 			& $0.140$ 	& $0.130$ 		& $0.33$ 	\\
      $	K $ 			& $0.494$ 	& $0.160$ 		& $0.32$ 	\\
      $\eta$  			& $0.548$ 	& $\cdots$		& $\cdots$	\\
      $\eta^\prime$  	& $0.958$ 	& $\cdots$		& $\cdots$	\\
      $\eta_q$  		& $\cdots$	& $0.139$ 		& $0.20$ \\
      $\eta_s$  		& $\cdots$  & $0.174$ 		& $0.34$ \\
      $\rho$ 			& $0.775$ 	& $0.216/0.165$ & $0.15$ \\
      $K^*$ 			& $0.892$ 	& $0.220/0.185$ & $0.17$ \\
      $\omega$ 			& $0.783$ 	& $0.187/0.151$ & $0.15$ \\
      $\phi$ 			& $1.019$ 	& $0.215/0.186$ & $0.19$ \\
    \end{tabular}
  \end{ruledtabular}
\end{table}

The cutoff scale $\mu_c$ is set at $1.0~\text{GeV}$, with contributions below the scale regarded as nonperturbative, corresponding to $\alpha_s/\pi > 0.2$. It has been verified that slight variations around this cutoff scale do not significantly affect the theoretical results \cite{Wang:2022ihx,Lu:2024jzn,Wang:2025jhq}. The soft contributions included in this work are soft form factors and color-octet parameters.

To determine the values of soft transition form factors, we employ the $B\to P(V)$ transition form factors either determined in theory of nonperturbative method or constrained by experimental data. The transition form factors in $B$-meson decays have been extensively investigated from both theoretical and experimental perspectives \cite{Ball:2004ye,Ball:2004rg,Ivanov:2007cw,Khodjamirian:2006st,Bailey:2015dka,Bharucha:2015bzk,Gubernari:2018wyi,Belle-II:2018jsg,Gao:2019lta,Parrott:2022rgu}. By employing the transition form factors $F_+$ and $A_0$ calculated through the LCSR approach and subtracting the perturbative contributions above the scale $\mu_c$, we extract the soft transition form factors for the relevant mesons as
\begin{align}
	\xi_+^{B\pi} 	&= 0.04\pm 0.01 ,	&\xi_{A0}^{B\rho}	&= 0.14\pm 0.04, 	\nonumber \\
	\xi_+^{BK} 		&= 0.04\pm 0.02 ,	&\xi_{A0}^{BK^*}	&= 0.16\pm 0.03,	\nonumber \\
	\xi_+^{B\eta_q} &= 0.06\pm 0.02 ,	&\xi_{A0}^{B\omega} &= 0.14\pm 0.04.
\end{align}
Given the quark composition of $\eta_s$ and $\phi$ mesons, both the soft and hard parts of the transition form factors for $B\to\eta_s$ and $B\to\phi$ are set to zero.

Regarding the soft production form factors and color-octet parameters, their inherent nonperturbative character obstructs the determination of these values through theoretical calculations. Consequently, these soft parameters are obtained by fitting experimental data. In this study, the chi-square analysis is utilized to estimate these parameters.
The chi-square statistic is defined as follows
\begin{equation}
	\chi^2=\sum_{i=1}^N\frac{(y_i-\mu_i)^2}{\sigma_i^2},
\end{equation}
where $\mu_i$ stands for the theoretical result for the $i$-th observable, $y_i$ the corresponding experimental central value, and $\sigma_i$ the standard deviation of experimental data. The summation over the index $i$ includes all branching ratios and $CP$-violation parameters for the relevant decay channels. Conventionally, the chi-square statistic is normalized by the degree of freedom (d.o.f) to yield the reduced chi-square statistic, defined as
\begin{equation}
\chi^2_{\text{d.o.f}}=\frac{\chi^2}{N-k},
\end{equation}
where $N$ denotes the number of experimental observables and $k$ the number of soft parameters to be determined.
By minimizing the $\chi^2_{\text{d.o.f}}$, the soft parameters are obtained. For the optimal fit, the reduced chi square is $\chi^2_{\text{d.o.f}}=62.4/(62-40)=2.8 $, and the corresponding parameters with the standard derivations are listed in Table~\ref{tab:fittedparam}.

The parameters $C_0^{a,b}$'s correspond to the color-octet contribution under SU(3) flavor symmetry, which stand for the magnitude of the color-octet contribution compared to color-singlet contributions. Table~\ref{tab:fittedparam} shows that the magnitudes of $C_0^{a}$ and $C_0^{b}$ are less than 0.2, which indicates that the contributions of the color-octet quark-antiquark pairs generated from the hard interaction are in general not large. Only for the decay mode where the contribution of the color-singlet state is suppressed, will the contribution of color-octet states may become significant. The values in Table~\ref{tab:fittedparam} also imply that the process of color-octet quark-antiquark pairs scattering into color-singlet meson states will generally generate nonzero phases. The parameter $C_0^{c}$ is relevant to the contribution of the final meson-pair production form factor within SU(3) symmetry, which originates from the factorizable annihilation diagrams [Figs. \ref{fig:eightdiagram}(g) and (h)]. In general, the contribution of the factorizable annihilation diagrams is small compared to that of Figs. \ref{fig:eightdiagram}(a) and (b). The parameters $R_i^j$'s stand for the SU(3) symmetry-breaking effects. The values of $R_i^j$ in Table~\ref{tab:fittedparam} indicate that the SU(3) symmetry-breaking effects are generally not small. One possible source of the SU(3) symmetry-breaking effects is the large mass differences between the final meson states.

We get a rather small value of the reduced chi-square $\chi^2_{\text{d.o.f}}$ in the numerical treating process, although some of the experimental data are not so precise. The reasons are: 1) The number of the experimental data used is large, totaling 62. This number substantially exceeds the number of free parameters. 2) There are experimental data whose precision is higher than 10\%, which is not so low. 3) To fit all the experimental data, although there are large uncertainties in a considerable portion of them, one has to narrow the theoretically allowed parameter region; otherwise, the theoretical output will overshoot the data for some decay modes, while undershooting the data for the others. Therefore, we get a fitting result with a small $\chi^2_{\text{d.o.f}}$.

\begin{table*}
  \caption{\label{tab:fittedparam} SU(3) parameters $C$ and $R$ for soft contributions are determined through fitting to experimental data. The columns denoted by ``$a$'', ``$b$'', and ``$c$'' represent the color-octet parameters for factorizable diagrams, nonfactorizable diagrams, and the parameters for production form factors, respectively. Note that the parameter $C_0^c$ corresponds to a quantity of dimension 1, whose unit is GeV here.
  }
  \begin{ruledtabular}
    \begin{tabular}{cccccccccccc}
    	& $a$ & $b$ & $c$ \\
		\hline
		$C_0$ 	& $(0.19 \pm 0.01 ) \mathrm{Exp}[( 0.53\pm 0.01) \pi i]$ & $(0.15 \pm 0.01 ) \mathrm{Exp}[( 0.74\pm 0.01) \pi i]$ & $(0.20 \pm 0.01 ) \mathrm{Exp}[( 0.80\pm 0.01) \pi i]$ \\
		$R_1^1$ & $(0.92 \pm 0.02 ) \mathrm{Exp}[( 0.79\pm 0.02) \pi i]$ & $(0.99 \pm 0.01 ) \mathrm{Exp}[( 0.62\pm 0.02) \pi i]$ & $(0.62 \pm 0.04 ) \mathrm{Exp}[(-0.59\pm 0.04) \pi i]$ \\
		$R_1^2$ & $(0.84 \pm 0.02 ) \mathrm{Exp}[( 0.82\pm 0.01) \pi i]$ & $(0.62 \pm 0.02 ) \mathrm{Exp}[( 1.00\pm 0.01) \pi i]$ & $(0.72 \pm 0.14 ) \mathrm{Exp}[( 0.89\pm 0.08) \pi i]$ \\
		$R_1^3$ & $\cdots$ 											     & $\cdots$												  & $(0.98 \pm 0.02 ) \mathrm{Exp}[( 0.87\pm 0.01) \pi i]$ \\
		$R_2^1$ & $(0.83 \pm 0.06 ) \mathrm{Exp}[( 0.16\pm 0.05) \pi i]$ & $(0.74 \pm 0.11 ) \mathrm{Exp}[(-0.27\pm 0.06) \pi i]$ & $(0.78 \pm 0.14 ) \mathrm{Exp}[( 0.25\pm 0.14) \pi i]$ \\
		$R_2^2$ & $(0.87 \pm 0.08 ) \mathrm{Exp}[(-0.11\pm 0.07) \pi i]$ & $(0.88 \pm 0.08 ) \mathrm{Exp}[(-0.29\pm 0.05) \pi i]$ & $\cdots$											   \\
		$R_2^3$ & $(0.79 \pm 0.10 ) \mathrm{Exp}[(-0.19\pm 0.07) \pi i]$ & $(0.83 \pm 0.07 ) \mathrm{Exp}[( 0.47\pm 0.04) \pi i]$ & $(0.76 \pm 0.14 ) \mathrm{Exp}[(-0.13\pm 0.12) \pi i]$ \\
		$R_2^4$ & $(0.79 \pm 0.08 ) \mathrm{Exp}[(-0.55\pm 0.06) \pi i]$ & $(0.43 \pm 0.12 ) \mathrm{Exp}[( 0.65\pm 0.07) \pi i]$ & $\cdots$											   \\
    \end{tabular}
  \end{ruledtabular}
\end{table*}

\begin{table*}
  \caption{\label{tab:btopvbr1} Branching ratios (in units of $10^{-6}$) for $B \to PV $ decay modes that are induced by the $b\to d$ transition. The experimental data \cite{ParticleDataGroup:2024cfk, Belle:2022dgi} and theoretical predictions from the PQCD\cite{Chai:2022ptk}, QCDF\cite{Sun:2014tfa,Cheng:2009cn}, SCET\cite{Wang:2008rk}, and FAT\cite{Zhou:2016jkv} frameworks are included for comparison. }
  \renewcommand\arraystretch{1.5}
  \begin{ruledtabular}
    \begin{tabular}{cccccccccccc}
		Mode & Exp. & This work & PQCD & QCDF & SCET & FAT \\
		\hline
		\multirow{2}{*}{$B^+\to \pi^0 \rho^+$}              & $10.6_{-1.3}^{+1.2}$            					& \multirow{2}{*}{$12.5_{-1.1-1.3}^{+1.5+2.1}$} & \multirow{2}{*}{$10.9_{-2.4-0.6}^{+3.4+0.6}$}		& \multirow{2}{*}{$10.9_{-0.8-2.4}^{+0.8+2.7}$}		& \multirow{2}{*}{$8.9_{-0.1-1.0}^{+0.3+1.0}$}	& \multirow{2}{*}{$12.9\pm 2.4$}	\\
															& $11.2\pm 1.1 _{-1.8}^{+1.2}$ \cite{Belle:2022dgi} &										 		&         											&         											&         										&         							\\
		$B^+\to \pi^+ \rho^0$                               & $8.3 \pm 1.2$                   					& $ 6.8_{-1.2-1.8}^{+1.5+1.9}  $ 				& $4.96_{-1.01-0.14}^{+1.34+0.13}$   				& $6.8_{-0.6-1.1}^{+0.6+1.2}$   					& $10.7_{-0.7-0.9}^{+0.7+1.0}$   				& $8.6\pm 2.3$						\\
		$B^0\to \pi^+ \rho^-+\pi^- \rho^+$                  & $23.0 \pm 2.3$                  					& $25.4_{-2.4-4.6}^{+2.9+3.7}  $ 				& $25.2_{-7.8-1.1}^{+5.7+1.1}$  					& $26.7_{-2.2-4.5}^{+2.1+5.1}$  					& $13.4_{-0.5-1.2}^{+0.6+1.2}$  				& $\cdots$							\\
		$B^0\to \pi^0 \rho^0$                               & $2.0 \pm 0.5$                   					& $ 2.6_{-0.7-1.1}^{+1.0+0.5}  $ 				& $0.32_{-0.07-0.04}^{+0.10+0.04}$  				& $1.2_{-0.1-0.5}^{+0.1+0.5}$  						& $2.5_{-0.1-0.2}^{+0.2+0.2}$  					& $1.32\pm 0.50$					\\ [0.5em]
		$B^+\to K^+ \overline{K}^{*0}$                      & $0.59 \pm 0.08$                 					& $ 0.51_{-0.10-0.19}^{+0.10+0.27} $ 			& $0.47_{-0.10-0.03}^{+0.15+0.02}$  				& $0.58_{-0.04-0.09}^{+0.03+0.09}$  				& $0.49_{-0.20-0.08}^{+0.26+0.09}$  			& $0.59\pm 0.12$  					\\
		$B^0\to K^{-} K^{*+}$                               & $<0.4$                          					& $ 0.13_{-0.03-0.08}^{+0.04+0.07} $ 			&\multirow{2}{*}{$0.39_{-0.03-0.03}^{+0.03+0.04}$}  & \multirow{2}{*}{$0.11_{-0.01-0.01}^{+0.01+0.01}$}	& $\cdots$										& $\cdots$							\\
		$B^0\to K^{+} K^{*-}$                               & $<0.4$                          					& $ 0.06_{-0.02-0.03}^{+0.02+0.03} $ 			&   												&   												& $\cdots$										& $\cdots$							\\
		$B^0\to K^0\overline{K}^{*0}+\overline{K}^0 K^{*0}$ & $<0.96$                         					& $ 0.90_{-0.11-0.50}^{+0.09+0.31} $ 			& $0.79_{-0.12-0.09}^{+0.17+0.11}$  				& $0.96_{-0.06-0.11}^{+0.05+0.13}$  				& $0.96_{-0.27-0.15}^{+0.34+0.18}$  			& $\cdots$							\\ [0.5em]
		$B^+\to \eta \rho^+$                                & $7.0 \pm 2.9$                   					& $ 9.5_{-1.2-2.5}^{+1.1+2.2}  $ 				& $5.59_{-1.22-1.06}^{+1.68+1.17}$   				& $8.3_{-0.6-0.9}^{+1.0+0.9}$   					& $3.9_{-1.7-0.4}^{+2.0+0.4}$   				& $8.16\pm 1.51$   					\\
		$B^+\to \eta^\prime \rho^+$                         & $9.7 \pm 2.2$                   					& $ 8.0_{-0.9-2.0}^{+1.2+1.0}  $ 				& $4.06_{-0.89-0.77}^{+1.22+0.84}$   				& $5.6_{-0.5-0.7}^{+0.9+0.8}$   					& $0.37_{-0.22-0.07}^{+2.46+0.08}$   			& $6.0\pm 1.0$   					\\
		$B^0\to \eta \rho^0$                                & $<1.5$                          					& $ 0.72_{-0.33-0.35}^{+0.31+0.39} $ 			& $0.13_{-0.02-0.02}^{+0.02+0.03}$ 					& $0.10_{-0.01-0.03}^{+0.02+0.04}$ 					& $0.04_{-0.01-0.00}^{+0.20+0.00}$ 				& $4.41\pm 1.23$ 					\\
		$B^0\to \eta^{\prime} \rho^0$                       & $<1.3$                          					& $ 1.56_{-0.28-0.72}^{+0.51+0.31} $ 			& $0.12_{-0.02-0.02}^{+0.03+0.02}$ 					& $0.09_{-0.04-0.03}^{+0.10+0.07}$ 					& $0.43_{-0.12-0.05}^{+2.51+0.05}$ 				& $3.91\pm 0.83$ 					\\ [0.5em]
		$B^+\to \pi^+ \omega$                               & $6.9 \pm 0.5$                   					& $ 6.3_{-1.2-1.6}^{+1.4+2.0}  $ 				& $5.42_{-1.10-0.45}^{+1.44+0.47}$   				& $6.7_{-1.0-1.1}^{+2.1+1.3}$   					& $6.7_{-0.3-0.6}^{+0.4+0.7}$   				& $6.78\pm 1.82$   					\\
		$B^0\to \pi^0 \omega$                               & $<0.5$                          					& $ 0.00_{-0.00-0.02}^{+0.05+0.08} $ 			& $0.10_{-0.01-0.01}^{+0.02+0.01}$  				& $0.01_{-0.00-0.01}^{+0.02+0.04}$  				& $0.0003_{-0.0000-0.0000}^{+0.0299+0.0000}$  	& $2.31\pm 0.91$  					\\
		$B^+\to \pi^+ \phi$                                 & $0.032 \pm 0.015$               					& $ 0.028_{-0.022-0.024}^{+0.041+0.025}$ 		& $0.042_{-0.010-0.002}^{+0.014+0.002}$ 			& $0.043$ 											& $0.003$ 										& $0.28\pm 0.06$ 					\\
		$B^0\to \pi^0 \phi$                                 & $<0.15$                         					& $ 0.01_{-0.01-0.01}^{+0.02+0.01} $ 			& $0.02_{-0.01-0.00}^{+0.01+0.00}$ 					& $0.01_{-0.01-0.01}^{+0.03+0.02}$ 					& $0.001$ 										& $0.13\pm 0.03$ 					\\
    \end{tabular}
  \end{ruledtabular}
\end{table*}
\begin{table*}
  \caption{\label{tab:btopvbr2} Branching ratios (in units of $10^{-6}$) for $B \to PV $ decay modes that are induced by the $b\to s$ transition. The experimental data \cite{ParticleDataGroup:2024cfk} and theoretical predictions from the PQCD\cite{Chai:2022ptk}, QCDF\cite{Sun:2014tfa,Cheng:2009cn}, SCET\cite{Wang:2008rk}, and FAT\cite{Zhou:2016jkv} frameworks are included for comparison. }
  \renewcommand\arraystretch{1.5}
  \begin{ruledtabular}
    \begin{tabular}{cccccccccccc}
		Mode & Exp. & This work & PQCD & QCDF & SCET & FAT \\
		\hline
		$B^+\to \pi^+ K^{*0}$        	& $10.1 \pm 0.8$     	& $ 10.0_{-1.9-1.5}^{+1.6+1.2} $	& $5.52_{-1.36-0.41}^{+1.93+0.38}$	& $8.7_{-0.5-1.2}^{+0.4+1.3}$   	& $8.5_{-3.6-1.4}^{+4.7+1.7}$ 	& $10.0\pm 2.0$ 	\\
		$B^+\to \pi^0 K^{*+}$        	& $6.8 \pm 0.9$      	& $ 5.7_{-0.8-0.7}^{+0.7+0.7}  $	& $3.58_{-0.82-0.15}^{+1.19+0.18}$	& $5.4_{-0.3-0.7}^{+0.3+0.7}$   	& $4.2_{-1.7-0.7}^{+2.2+0.8}$ 	& $6.23\pm 1.11$	\\
		$B^0\to \pi^- K^{*+}$        	& $7.5 \pm 0.4$      	& $ 8.0_{-1.2-1.0}^{+0.9+0.9}  $	& $4.67_{-1.09-0.27}^{+1.54+0.24}$	& $7.5_{-0.5-1.0}^{+0.4+1.1}$   	& $8.4_{-3.4-1.3}^{+4.4+1.6}$ 	& $8.40\pm 1.66$	\\
		$B^0\to \pi^0 K^{*0}$        	& $3.3 \pm 0.6$      	& $ 3.0_{-0.7-0.5}^{+0.5+0.5}  $	& $1.57_{-0.38-0.16}^{+0.55+0.14}$	& $2.9_{-0.2-0.5}^{+0.1+0.5}$   	& $4.6_{-1.8-0.7}^{+2.3+0.9}$ 	& $3.35\pm 0.75$	\\ [0.5em]
		$B^+\to K^0 \rho^+$          	& $7.3_{-1.2}^{+1.0}$	& $ 8.5_{-2.2-1.4}^{+0.6+1.5}  $	& $6.11_{-0.34-0.86}^{+0.45+0.96}$	& $7.9_{-0.5-1.1}^{+0.4+1.3}$   	& $9.3_{-3.7-1.4}^{+4.7+1.7}$ 	& $7.74\pm 1.62$	\\
		$B^+\to K^+ \rho^0$          	& $3.7 \pm 0.5$      	& $ 2.7_{-0.8-0.6}^{+0.5+0.6}  $	& $3.28_{-0.21-0.48}^{+0.25+0.50}$	& $3.41_{-0.21-0.57}^{+0.19+0.63}$ 	& $6.7_{-2.2-0.9}^{+2.7+1.0}$ 	& $3.97\pm 0.84$	\\
		$B^0\to K^+ \rho^-$          	& $7.0 \pm 0.9$      	& $ 6.3_{-1.3-0.9}^{+0.6+1.4}  $	& $6.05_{-0.42-0.75}^{+0.57+0.84}$	& $9.0_{-0.5-1.3}^{+0.5+1.4}$   	& $9.8_{-3.7-1.4}^{+4.6+1.7}$ 	& $8.27\pm 1.71$	\\
		$B^0\to K^0 \rho^0$          	& $3.4 \pm 1.1$      	& $ 4.6_{-1.1-0.6}^{+0.4+1.2}  $	& $3.64_{-0.38-0.45}^{+0.49+0.47}$	& $5.5_{-0.3-0.7}^{+0.3+0.8}$   	& $3.5_{-1.5-0.6}^{+2.0+0.7}$ 	& $4.59\pm 0.86$	\\ [0.5em]
		$B^+\to \eta K^{*+}$         	& $19.3 \pm 1.6$     	& $17.7_{-4.2-2.2}^{+2.4+2.5}  $	& $6.08_{-0.30-1.45}^{+0.41+2.02}$	& $15.7_{-4.3-7.1}^{+8.5+9.4}$  	& $17.9_{-5.4-2.9}^{+5.5+3.5}$	& $17.3\pm 2.5$ 	\\
		$B^+\to \eta^{\prime} K^{*+}$	& $4.8_{-1.6}^{+1.8}$	& $ 3.7_{-1.9-1.7}^{+2.6+2.7}  $	& $1.54_{-0.34-0.08}^{+0.51+0.17}$	& $1.7_{-0.4-1.6}^{+2.7+4.1}$   	& $4.5_{-3.9-0.8}^{+6.6+0.9}$ 	& $3.31\pm 0.60$	\\
		$B^0\to \eta K^{*0}$         	& $15.9 \pm 1.0$     	& $17.0_{-4.1-2.3}^{+2.8+1.9}  $	& $5.37_{-0.25-1.25}^{+0.35+1.69}$	& $15.6_{-4.1-7.1}^{+7.9+9.4}$   	& $16.6_{-5.0-2.7}^{+5.1+3.2}$	& $16.6\pm 2.4$ 	\\
		$B^0\to \eta^{\prime} K^{*0}$	& $2.8 \pm 0.6$      	& $ 2.5_{-1.4-1.1}^{+1.6+1.5}  $	& $1.60_{-0.32-0.03}^{+0.46+0.04}$	& $1.5_{-0.4-1.7}^{+2.4+3.9}$   	& $4.1_{-3.6-0.7}^{+6.2+0.9}$ 	& $3.0\pm 0.6$  	\\ [0.5em]
		$B^+\to K^+ \omega$          	& $6.5 \pm 0.4$      	& $ 6.5_{-2.8-0.8}^{+5.0+0.7}  $	& $6.17_{-0.90-1.33}^{+1.25+1.59}$	& $4.8_{-1.9-2.3}^{+4.4+3.5}$   	& $5.1_{-1.9-0.8}^{+2.4+0.9}$ 	& $6.52\pm 1.35$	\\
		$B^0\to K^0 \omega$          	& $4.8 \pm 0.4$      	& $ 4.7_{-1.6-0.7}^{+3.2+0.8}  $	& $5.60_{-0.77-1.35}^{+0.99+1.55}$	& $4.1_{-1.7-2.2}^{+4.2+3.3}$   	& $4.1_{-1.7-0.7}^{+2.1+0.8}$ 	& $4.80\pm 1.13$	\\
		$B^+\to K^+ \phi$            	& $8.8_{-0.6}^{+0.7}$	& $ 8.2_{-3.4-1.8}^{+4.1+1.9}  $	& $4.61_{-0.82-0.63}^{+1.41+2.29}$	& $8.8_{-2.7-3.6}^{+2.8+4.7}$   	& $9.7_{-3.9-1.5}^{+4.9+1.8}$ 	& $8.38\pm 1.48$	\\
		$B^0\to K^0 \phi$            	& $7.3 \pm 0.7$      	& $ 7.6_{-3.1-1.6}^{+3.8+1.8}  $	& $4.25_{-0.71-0.55}^{+1.25+0.66}$	& $8.1_{-2.5-3.3}^{+2.6+4.4}$   	& $9.1_{-3.6-1.4}^{+4.6+1.7}$ 	& $7.77\pm 1.37$	\\ 
    \end{tabular}
  \end{ruledtabular}
\end{table*}
\begin{table*}
  \caption{\label{tab:btopvcpv1} $CP$ asymmetries (in units of $10^{-2}$) for $B \to PV $ decay modes that are induced by the $b\to d$ transition. The experimental data \cite{ParticleDataGroup:2024cfk, BaBar:2013uwm} and theoretical predictions from the PQCD\cite{Chai:2022ptk}, QCDF\cite{Sun:2014tfa,Cheng:2009cn}, SCET\cite{Wang:2008rk}, and FAT\cite{Zhou:2016jkv} frameworks are included for comparison.}
  \renewcommand\arraystretch{1.5}
  \begin{ruledtabular}
    \begin{tabular}{cccccccccccc}
      	Mode & Exp. & This work & PQCD & QCDF & SCET & FAT \\
      	\hline
      	$B^+\to \pi^0 \rho^+$                 	& $3 \pm 10 $									& $ 8.1_{-5.6-3.9}^{+3.3+3.2} $ 				& $-7.31_{-0.02-0.03}^{+0.06+0.07}$					& $8.2_{-0.3-1.5}^{+0.2+1.6}$  		& $15.5_{-18.9-1.4}^{+16.9+1.6}$  					& $16\pm 2$						\\
      	$B^+\to \pi^+ \rho^0$                 	& $0.3\pm 1.4$      							& $-1.1_{-5.5-3.8}^{+4.8+5.1}$ 					& $14.9_{-0.4-0.6}^{+0.4+0.5}$ 						& $-6.7_{-0.2-3.7}^{+0.2+3.2}$ 		& $-10.8_{-12.7-0.7}^{+13.1+0.9}$ 					& $-45\pm 4$ 					\\
      	\multirow{2}{*}{$B^0\to \pi^+ \rho^-$}	& $-8 \pm 8$        							& \multirow{2}{*}{$-21.9_{-3.8-3.9}^{+4.4+6.2}$}& \multirow{2}{*}{$-30.9_{-0.1-1.6}^{+0.1+1.7}$}	& \multirow{2}{*}{$\cdots$} 		& \multirow{2}{*}{$-9.9_{-16.7-0.7}^{+17.2+0.9}$} 	& \multirow{2}{*}{$-44\pm 3$} 	\\
		      									& $-12\pm 8 _{-5}^{+4}$ \cite{BaBar:2013uwm}	&              									&         											&		  							&		    										&		  						\\
      	\multirow{2}{*}{$B^0\to \pi^- \rho^+$}	& $13 \pm 6$         							& \multirow{2}{*}{$-0.9_{-2.3-3.3}^{+2.0+2.6}$}	& \multirow{2}{*}{$-0.59_{-0.17-0.68}^{+0.18+0.72}$}& \multirow{2}{*}{$\cdots$}  		& \multirow{2}{*}{$11.8_{-20.0-1.1}^{+17.5+1.2}$}	& \multirow{2}{*}{$15\pm 3$}  	\\
												& $ 9_{-6}^{+5} \pm 4$ \cite{BaBar:2013uwm} 	&              									&         											&         							&         											&         						\\
      	$C_{B^0\to \pi^0 \rho^0}$             	& $27 \pm 24$         							& $ 2.0_{-3.5-16.1}^{+4.7+5.4} $ 				& $66.1_{-1.7-3.6}^{+2.1+3.9}$ 						& $3.9_{-0.1-5.0}^{+0.1+5.1}$ 		& $0.6_{-21.4-0.1}^{+21.9+0.1}$ 					& $-36\pm 8$ 					\\
      	$S_{B^0\to \pi^0 \rho^0}$             	& $-23 \pm 34$        							& $ 12.6_{-6.1-7.0}^{+4.8+4.8} $ 				& $-46.8_{-0.8-3.5}^{+0.1+2.7}$ 					& $-29_{-7-5}^{+5+3}$ 				& $\cdots$											& $19\pm 16$ 					\\ [0.5em]
		$B^+\to K^+ \overline{K}^{*0}$        	& $4 \pm 5$         							& $ 0.1_{-8.1-13.8}^{+3.5+9.6} $ 				& $21.3_{-5.7-1.4}^{+6.2+1.2}$ 						& $-10.6_{-0.4-2.6}^{+0.3+3.0}$		& $-3.6_{-5.3-0.4}^{+6.1+0.4}$ 						& $-10\pm 2$ 					\\
		$B^+\to \eta \rho^+$                  	& $11 \pm 11$         							& $ 1.8_{-7.5-10.1}^{+12.8+13.6} $ 				& $-13.0_{-0.1-1.5}^{+0.1+0.1}$ 					& $-8.5_{-0.4-5.3}^{+0.4+6.5}$ 		& $-6.6_{-21.3-0.7}^{+21.5+0.6}$ 					& $-11\pm 2$ 					\\
		$B^+\to \eta^\prime \rho^+$           	& $26 \pm 17$         							& $-20.1_{-8.8-14.7}^{+8.8+18.5} $ 				& $29.0_{-0.4-0.1}^{+0.4+0.0}$  					& $1.4_{-2.2-11.7}^{+0.8+14.0}$  	& $-19.8_{-37.5-3.1}^{+66.5+2.8}$  					& $45\pm 5$  					\\
		$B^+\to \pi^+ \omega$                 	& $-4 \pm 5$        							& $-3.4_{-4.8-8.4}^{+5.1+6.4} $ 				& $-29.8_{-0.4-0.8}^{+0.5+1.1}$ 					& $-13.2_{-2.1-10.7}^{+3.2+12.0}$ 	& $0.5_{-19.6-0.0}^{+19.1+0.1}$ 					& $5.4\pm 5.2$ 					\\
		$B^+\to \pi^+ \phi$                   	& $10 \pm 50$        							& $ 0.0_{-0.1-0.2}^{+0.2+0.3}  $ 				& $0.0$  											& $0.0$  							& $\cdots$											& $\cdots$						\\
    \end{tabular}
  \end{ruledtabular}
\end{table*}
\begin{table*}
  \caption{\label{tab:btopvcpv2} $CP$ asymmetries (in units of $10^{-2}$) for $B \to PV $ decay modes that are induced by the $b\to s$ transition. The experimental data \cite{ParticleDataGroup:2024cfk} and theoretical predictions from the PQCD \cite{Chai:2022ptk}, QCDF \cite{Sun:2014tfa,Cheng:2009cn}, SCET \cite{Wang:2008rk}, and FAT\cite{Zhou:2016jkv} frameworks are included for comparison.}
  \renewcommand\arraystretch{1.5}
  \begin{ruledtabular}
    \begin{tabular}{cccccccccccc}
      	Mode & Exp. & This work & PQCD & QCDF & SCET & FAT \\
      	\hline
      	$B^+\to \pi^+ K^{*0}$        	& $-2.1 \pm 3.2$  	& $-0.1_{-0.1-0.1}^{+0.3+0.2}$ 		& $-0.94_{-0.29-0.03}^{+0.26+0.04}$ 	& $0.47_{-0.02-0.13}^{+0.02+0.11}$	& $0$ 								& $0.5\pm 0.1$ 	\\
      	$B^+\to \pi^0 K^{*+}$        	& $-39 \pm 21$    	& $-5.7_{-1.0-6.8}^{+1.0+8.1} $ 	& $-0.01_{-4.87-1.26}^{+4.40+1.12}$ 	& $0.4_{-0.0-4.7}^{+0.0+4.0}$ 		& $-17.8_{-24.6-2.0}^{+30.3+2.2}$ 	& $8.8\pm 4.0$ 	\\
      	$B^0\to \pi^- K^{*+}$        	& $-27 \pm 4$     	& $-26.8_{-5.6-6.0}^{+4.7+6.8} $ 	& $-14.1_{-6.4-3.1}^{+6.0+2.9}$ 		& $-26_{-1-1}^{+1+1}$ 				& $-11.2_{-16.2-1.3}^{+19.0+1.3}$ 	& $-20\pm 4$ 	\\
      	$B^0\to \pi^0 K^{*0}$        	& $-15 \pm 13$    	& $-18.1_{-5.8-7.8}^{+4.6+8.0} $ 	& $-14.8_{-0.1-1.5}^{+0.4+1.2}$ 		& $-21_{-1-6}^{+1+6}$ 				& $5.0_{-8.4-0.5}^{+7.5+0.5}$ 		& $-27\pm 5$ 	\\ [0.5em]
      	$B^+\to K^0 \rho^+$          	& $-3 \pm 15$     	& $ 1.9_{-0.4-0.3}^{+2.0+0.2} $ 	& $0.99_{-0.01-0.18}^{+0.01+0.13}$  	& $1.3_{-0.1-0.1}^{+0.1+0.1}$  		& $0$  								& $0.9\pm 0.0$  \\
      	$B^+\to K^+ \rho^0$          	& $16.0 \pm 2.1$  	& $ 15.0_{-6.4-5.0}^{+14.0+4.9}$	& $58.7_{-4.0-2.8}^{+4.3+3.2}$  		& $26_{-1-5}^{+1+5}$  				& $9.2_{-16.1-0.7}^{+15.2+0.7}$  	& $59\pm 6$  	\\
      	$B^0\to K^+ \rho^-$          	& $20 \pm 11$     	& $ 24.8_{-4.8-12.8}^{+8.5+7.2} $ 	& $54.3_{-0.4-0.7}^{+0.6+0.8}$  		& $27_{-1-3}^{+1+3}$  				& $7.1_{-12.4-0.7}^{+11.2+0.7}$  	& $59\pm 1$  	\\
      	$C_{B^0\to K^0_S \rho^0}$    	& $-4 \pm 20$     	& $-17.9_{-7.1-6.1}^{+7.0+15.1} $ 	& $8.96_{-1.48-0.01}^{+1.81+0.05}$ 		& $-15_{-1-3}^{+1+3}$ 				& $6.6_{-11.6-0.8}^{+9.7+0.9}$ 		& $8.5\pm 5.9$ 	\\
      	$S_{B^0\to K^0_S \rho^0}$    	& $50_{-21}^{+17}$	& $ 66.8_{-3.8-8.3}^{+3.5+4.8} $ 	& $57.9_{-0.4-0.0}^{+0.5+0.1}$  		& $63_{-2-2}^{+2+3}$  				& $\cdots$							& $88\pm 5$  	\\ [0.5em]
		$B^+\to \eta K^{*+}$         	& $2 \pm 6$       	& $-5.2_{-1.6-4.2}^{+1.9+3.8} $ 	& $-34.5_{-2.4-0.8}^{+2.5+0.9}$ 		& $-9.7_{-3.7-7.1}^{+3.9+6.2}$ 		& $-2.6_{-5.5-0.3}^{+5.4+0.3}$ 		& $-17\pm 2$ 	\\
		$B^+\to \eta^{\prime} K^{*+}$	& $-26 \pm 27$    	& $ 0.7_{-7.9-16.2}^{+10.3+13.9} $ 	& $1.54_{-8.16-9.74}^{+9.05+14.9}$ 		& $65.5_{-39.5-50.2}^{+10.1+34.2}$ 	& $2.7_{-19.5-0.3}^{+27.4+0.4}$ 	& $-45\pm 9$ 	\\
		$B^0\to \eta K^{*0}$         	& $19 \pm 5$      	& $ 7.5_{-2.3-3.0}^{+2.6+2.5} $ 	& $2.10_{-0.55-0.21}^{+0.71+0.18}$  	& $3.5_{-0.5-2.4}^{+0.4+2.7}$  		& $-1.1_{-2.4-0.1}^{+2.3+0.1}$  	& $6.5\pm 1.1$  \\
		$B^0\to \eta^{\prime} K^{*0}$	& $-7 \pm 18$     	& $-8.7_{-17.4-14.7}^{+12.1+23.3} $ & $12.4_{-0.3-1.7}^{+0.1+0.5}$ 			& $6.8_{-9.2-50.2}^{+10.7+33.2}$ 	& $9.6_{-11.0-1.2}^{+8.9+1.3}$ 		& $5.9\pm 4.9$ 	\\ [0.5em]
		$B^+\to K^+ \omega$          	& $-2 \pm 4$      	& $-3.7_{-5.2-6.7}^{+11.5+7.6} $ 	& $31.5_{-1.1-0.7}^{+0.6+0.1}$  		& $22.1_{-12.8-13.0}^{+13.7+14.0}$ 	& $11.6_{-20.4-1.1}^{+18.2+1.1}$  	& $19\pm 9$  	\\
		$C_{B^0\to K^0_S \omega}$    	& $0 \pm 40$      	& $-16.1_{-2.4-5.0}^{+6.8+12.4}  $ 	& $-5.29_{-0.99-0.40}^{+0.83+0.21}$ 	& $4.7_{-1.8-5.5}^{+1.6+5.8}$ 		& $-5.2_{-8.0-0.6}^{+9.2+0.6}$	 	& $-25\pm 10$ 	\\
		$S_{B^0\to K^0_S \omega}$    	& $70 \pm 21$     	& $ 68.6_{-5.5-8.0}^{+5.7+5.3} $ 	& $79.2_{-0.2-0.2}^{+0.1+0.3}$  		& $84_{-5-6}^{+5+4}$  				& $\cdots$							& $70\pm 4$  	\\
		$B^+\to K^+ \phi$            	& $1.7 \pm 1.7$   	& $-0.3_{-0.1-0.5}^{+0.2+1.1}$ 		& $-1.93_{-0.60-0.42}^{+0.66+0.66}$ 	& $0.6_{-0.1-0.1}^{+0.1+0.1}$ 		& $0$ 								& $-0.6\pm 0.1$ \\
		$C_{B^0\to K^0_S \phi}$      	& $-9 \pm 12$     	& $ 1.5_{-0.2-1.8}^{+0.3+1.1} $ 	& $2.67_{-0.18-0.26}^{+0.05+0.28}$ 		& $-0.9_{-0.2-0.2}^{+0.1+0.1}$ 		& $0$ 								& $0.6\pm 0.1$ 	\\
		$S_{B^0\to K^0_S \phi}$      	& $58 \pm 12$     	& $ 71.3_{-0.1-1.3}^{+0.1+0.7} $ 	& $70.6_{-0.3-0.1}^{+0.3+0.8}$  		& $69.2_{-0.0-0.2}^{+0.3+0.2}$  	& $\cdots$							& $70\pm 0$  	\\
    \end{tabular}
  \end{ruledtabular}
\end{table*}

Including the soft contributions, the theoretical results for branching ratios and $CP$ violations in charmless $B\to PV$ decays are summarized in Tables.~\ref{tab:btopvbr1}, \ref{tab:btopvbr2}, \ref{tab:btopvcpv1} and \ref{tab:btopvcpv2}. For each decay mode, we first list the averaged experimental data compiled by the Particle Data Group \cite{ParticleDataGroup:2024cfk}, labeled as ``Exp.''. Regarding the $\pi^\pm \rho^\mp$ and $K^0\overline{K}^{*0}(\overline{K}^0K^{*0})$ channels, the $B^0-\bar{B}^0$ mixing makes it challenging to distinguish these related channels in experiments. Thus, the branching ratios presented in these tables are summed over these channels. Following these experimental data, we calculate the same quantities as experimental measurement, and the results of our calculations, denoted as ``This work'', are provided. The uncertainties are mainly caused by the uncertainties of the parameters in mesons' distribution amplitudes and the parameters for  the soft contributions, which are listed as the first and second uncertainties, respectively. We also include the results calculated from the ``PQCD''\cite{Chai:2022ptk}, ``QCDF''\cite{Sun:2014tfa,Cheng:2009cn}, ``SCET''\cite{Wang:2008rk}, and ``FAT''\cite{Zhou:2016jkv} approaches within these tables for comparison.

Note that the sources of the uncertainties of the PQCD approach used in Ref. \cite{Chai:2022ptk} and our modified PQCD approach are very different, although these two approaches are very similar in the perturbative part. One of the main uncertainties comes from the $B$ meson wave function. For the perturbative part, the $B$ meson wave function used in our approach is different from the one used in Ref. \cite{Chai:2022ptk}, which causes different uncertainties. Second, the uncertainty of the $B$ meson wave function only affects the perturbative part. It does not affect the soft contributions from the soft form factors in this work. Therefore, the uncertainty structure in our approach is very different from that of Ref. \cite{Chai:2022ptk}.

In Tables.~\ref{tab:btopvbr1} and \ref{tab:btopvbr2}, we find that the branching ratios calculated in this work are well consistent with the experimental measurements for almost all the decay channels. For the $B^+\to\pi^0\rho^+$ decay mode, our result for the branching ratio agrees preferably with the recent measurement by Belle collaboration \cite{Belle:2022dgi}. Moreover, the calculation in this work shows the most favorable agreement with experimental data in $\eta^\prime\rho^+$ decay mode compared with all the other theoretical methods. However, given that large uncertainties exist in both the theoretical results and experimental data for this decay mode, the agreement may not be so significant. More precise experimental data and theoretical calculations are necessary.

\begin{table}
  \caption{\label{tab:brpart}
  Branching ratios (in units of $10^{-6}$) for several selected decay modes. The label ``$\text{LO}_{\text{NLOWC}}$'' means the contribution of LO diagrams with the NLO Wilson coefficients, ``NLO'' the result with the inclusion of all the NLO diagrams, ``$+\xi^{BP(V)}$'' the inclusion of the soft transition form factors based on NLO,  ``$+\xi^{PV}$'' the inclusion of the soft production form factors based on NLO+$\xi^{BP(V)}$, and ``+CO'' the inclusion of the color-octet contributions based on NLO+$\xi^{BP(V)}+\xi^{PV}$.}
  \begin{ruledtabular}
    \begin{tabular}{cccccccccccc}
		Mode & $\text{LO}_{\text{NLOWC}}$ & NLO & $+\xi^{BP(V)}$ & $+\xi^{PV}$ & +CO \\
		\hline
		$\pi^0\rho^0$ 		& $0.24$ & $0.05$ & $0.19$ & $0.19$ & $2.60$ \\
		$\eta^\prime\rho^+$ & $2.1$  & $1.8$  & $2.2$  & $2.2$  & $8.0$  \\
		$\pi^0 K^{*+}$ 		& $4.1$  & $3.1$  & $4.9$  & $6.2$  & $5.7$  \\
    \end{tabular}
  \end{ruledtabular}
\end{table}

To investigate the different contributions of each theoretical component to the decay amplitudes, we present the branching ratios for several decay modes in Table.~\ref{tab:brpart}. The large difference between the contributions of NLO and LO diagrams in the $B^0\to \pi^0\rho^0$ decay mode is caused by the smallness of the LO contribution, where cancellation seriously occurs for the LO contributions due to the isospin structure of $\pi^0\rho^0$. For the other decay modes without such cancellation in the LO contributions, the contributions of NLO diagrams are only about 10\% to 20\%. In the $\pi^0\rho^0$ and $\eta^\prime\rho^+$ decay channels, the contributions from the annihilation diagrams, and consequently from the soft production form factors, strongly cancel out. So the values in the column `$+\xi^{PV}$'' remain the same as those in the column ``$+\xi^{BP(V)}$'' for these two decay modes. The branching ratios of these two channels are dramatically enhanced only by the color-octet contributions, which can be seen from the comparison of the last two columns. In contrast, the situation is different for the $\pi^0 K^{*+}$ decay mode, as both the soft transition and production form factors improve the results significantly, which has already made them comparable to the experimental measurements. Note that the values of LO+NLO (the column NLO in Table~\ref{tab:brpart}) are still not the same as those given by PQCD in Ref. \cite{Chai:2022ptk}, because a new $B$ meson wave function is used in this work and an infrared cutoff scale is introduced, which are different from Ref. \cite{Chai:2022ptk}.

In Tables.~\ref{tab:btopvcpv1} and \ref{tab:btopvcpv2}, we present the results of direct $CP$ violations and mixing-induced $CP$ violations for the decay channels that have been experimentally measured. The theoretical calculations in this work are generally consistent with the experimental data within the experimental and theoretical uncertainties. For the $\pi^\pm\rho^\mp$ decays, the direct $CP$ violations obtained in this work are more closely in alignment with the measurement of the BaBar collaboration \cite{BaBar:2013uwm} than with the averaged experimental result. There are several channels, including the $\pi^0\rho^0$, $\eta^\prime\rho^+$, $\pi^0K^{*+}$, and $\eta^\prime K^{*+}$ decay modes, where the theoretical output presented in this paper seems to deviate considerably from the experimental data. However, it is important to note that the experimental measurements in these decay channels also suffer from significant uncertainties, which makes it unfeasible to confirm the validity of different theoretical frameworks. Hence, more precise data are necessary for further investigations.

Finally, we would like to make a brief remark about the vector mesons. The vector mesons usually appear as short-lived resonances in experiments, and are reconstructed from their decay daughter particles. They are usually described by a Breit-Wigner ansatz with resonance parameters, such as mass and decay width, incorporated in it. Here, we treat the vector mesons as stable particles without considering decay width effects in this work.

\section{Summary \label{sec:sum}}
In this work we present a phenomenological analysis of the charmless $B$-meson decays, with the final states consisting of a light pseudoscalar and a light vector meson. Based on the NLO PQCD calculations, we introduce the soft form factors and color-octet parameters. They serve to absorb the nonperturbative contributions below the infrared cutoff scale $\mu_c$ and the effect of color-octet contributions. By employing the SU(3) symmetry and its breaking effects for quark flavors, these parameters can be expressed in terms of a more compact set of SU(3) parameters. The implementation of the SU(3) symmetry significantly reduces the number of parameters and preserves the predictive power of the theoretical framework. To estimate the values of the SU(3) parameters, we perform a $\chi^2$ fitting procedure using the experimental data. Compared with the $B\to PP$ decays, the $B\to PV$ decays involve a greater number of decay channels and observables, thereby imposing stricter constraints on the parameters.

With the fitted parameters, the theoretical output for branching ratios and $CP$ violations can be obtained. We conclude that most experimental measurements of branching ratios and $CP$ asymmetries can be satisfactorily explained within this framework. Soft form factors and color-octet contributions play an essential role especially in decay channels that cannot be well described previously, such as $\pi^0\rho^0$ and $\eta^\prime\rho^+$ decay modes. The analysis for the $CP$ asymmetries is hampered by the large uncertainties, which requires more precise experimental data.


\begin{acknowledgments}
	This work is supported in part by the National Natural Science Foundation of China under Contracts No. 12275139, No. 12535006, and No. 11875168, and S.L. is partially supported by the  Postdoctoral Fellowship Program of CPSF under Grant No. GZC20252229 and supported by the Postdoctoral Science Foundation under Grant No. 2025M773381.
\end{acknowledgments}

\appendix
\section{\uppercase{The wave function of $B$ meson} \label{app:bwavefunc}}
Considering the heavy mass of the $b$ quark compared to the $u$ and $d$ quarks, the $B$ meson is viewed as a heavy-light two-body system. The wave function of the $B$ meson is defined through the nonlocal matrix element as
\begin{equation}
	\langle 0|\bar{q}_{\gamma}(z)[z,0]b_{\delta}(0)|\bar{B}(\vec{k})\rangle = \int d^3k \Phi_{\delta\gamma}^B(\vec{k},t)\exp(-ik\cdot z),
\end{equation}
where $\gamma$ and $\delta$ represent spinor indices, and the symbol $[z,0]$ represents the Wilson line, which can preserve the gauge invariance of the wave function. Rather than the form given in Ref.~\cite{Beneke:2000wa}, we employ an alternative formulation of the matrix element, which originates from the QCD-inspired relativistic potential model \cite{Yang:2011ie,Liu:2013maa,Liu:2015lka,Sun:2016avp}. The spinor wave function is expressed as follows \cite{Sun:2016avp}
\begin{align} \label{eq:PhiB}
	\Phi_{\delta\gamma}^B(\vec{k})&=\frac{-if_Bm_B}{4}K(\vec{k}) \nonumber \\
		&\times\biggl\{(E_Q+m_Q)\frac{1+\not{v}}{2}\biggl[\left(\frac{k_+}{\sqrt{2}}+\frac{m_q}{2}\right)\not{n}_+ \nonumber \\
		&\quad+\left(\frac{k_-}{\sqrt{2}}+\frac{m_q}{2}\right)\not{n}_{-}-k_\perp^\mu\gamma_\mu\biggr]\gamma_5 \nonumber \\
		&\quad-(E_q+m_q)\frac{1-\not{v}}{2}\biggl[\left(\frac{k_+}{\sqrt{2}}-\frac{m_q}{2}\right)\not{n}_+ \nonumber \\
		&\quad+\left(\frac{k_-}{\sqrt{2}}-\frac{m_q}{2}\right)\not{n}_--k_\perp^\mu\gamma_\mu\biggr]\gamma_5\biggr\}_{\delta\gamma},
\end{align}
where $v$ is the four velocity of the $B$ meson. The subscripts $Q$ and $q$, attached to the energies $E$ and masses $m$, distinguish the heavy and light quark components of the $B$ meson. Within the light-cone coordinate, the lightlike vectors $n_{\pm}$ in the expression are defined as $n_{\pm}^\mu = (1, 0, 0, \mp 1)$. The momentum $k$ associated with the light quark is decomposed in the light-cone coordinate as $k^\pm=(E_q\pm k^3)/\sqrt{2}$ and $k_\perp^\mu=(0,k^1,k^2,0)$. The function $K(\vec{k})$, which is directly proportional to the momentum-space wave function $\Psi_0(\vec{k})$, is given by
\begin{align}
	K(\vec{k}) &= \frac{2N_B\Psi_0(\vec{k})}{\sqrt{E_qE_Q(E_q+m_q)(E_Q+m_Q)}} \nonumber \\
		&= \frac{2N_Ba_1 \text{Exp}\left(a_2 |\vec{k}|^2 + a_3 |\vec{k}| +a_4\right)}{\sqrt{E_qE_Q(E_q+m_q)(E_Q+m_Q)}},
\end{align}
where $N_B=\sqrt{{3}/\left[(2\pi)^3m_B\right]}/f_B$ is the normalization factor. The related parameters $a_{1-4}$ are obtained from the numerical results of the relativistic potential model. Their values have been obtained in Ref.~\cite{Sun:2016avp}, which are
\begin{align} \label{eq:a1234}
	a_1 &= 4.55_{-0.30}^{+0.40}~\text{GeV}^{-3/2}, & a_2 &= -0.39_{-0.20}^{+0.15}~\text{GeV}^{-2}, \nonumber\\
    a_3 &= -1.55\pm 0.20~\text{GeV}^{-1}, & a_4 &= -1.10_{-0.05}^{+0.10}.
\end{align}

\begin{widetext}
\section{\uppercase{The Expressions for $B\to PV$ diagrams} \label{app:fullamp}}
For $B\to P_1V_2$ transitions, the amplitudes corresponding to the four types of LO topological diagrams and NLO corrections are summarized below
{
\setlength{\jot}{8pt}
\begin{align} \label{eq:me}
	\mathcal{M}_{e,P_1V_2}=&16\pi^2C_Fm_B^2f_B
		\int k_{0\perp}dk_{0\perp}\int dx_0 dx_1 dx_2 \int b_0db_0 b_2db_2
		G_B(x_0,b_0,k_{0\perp}) \phi_{V_2}(x_2,b_2) \{\alpha_s(t_{ne}) [-x_2 \nonumber \\
	&\times (E_q-k_0^z)\phi_{P_1}^A(x_1,b_0)+r_{P_1}(1-x_1)(E_q+k_0^z)(\phi_{P_1}^P(x_1,b_0)+\phi_{P_1}^T(x_1,b_0))]
		h_{ne}(x_0,x_2,1-x_1,b_0,b_2) \nonumber \\
	&\times \exp[-S_B(t_{ne})-S_{P_1}(t_{ne})|_{b_1\to b_0}-S_{V_2}(t_{ne})]+\alpha_s(t_{ne}^\prime)
		[((2-x_1-x_2)E_q-(x_1-x_2)k_0^z)\phi_{P_1}^A(x_1,b_0) \nonumber \\
	&-r_{P_1}(1-x_1)(E_q-k_0^z)(\phi_{P_1}^P(x_1,b_0)-\phi_{P_1}^T(x_1,b_0))]
		\exp[-S_B(t_{ne}^\prime)-S_{P_1}(t_{ne}^\prime)|_{b_1\to b_0}-S_{V_2}(t_{ne}^\prime)]\nonumber \\
	&\times h_{ne}(x_0,1-x_2,1-x_1,b_0,b_2)\},
\end{align}
\begin{align} \label{eq:mer}
	\mathcal{M}_{e,P_1 V_2}^R=&16\pi^2C_Fm_B^2f_B
		\int k_{0\perp}dk_{0\perp} \int dx_0 dx_1 dx_2 \int b_0db_0 b_2db_2
		G_B(x_0,b_0,k_{0\perp}) (-r_{V_2})\{\alpha_s(t_{ne})[x_2(E_q-k_0^z) \nonumber \\
	&\times \phi_{P_1}^A(x_1,b_0)(-\phi_{V_2}^s(x_2,b_2)+\phi_{V_2}^t(x_2,b_2))
		+r_{P_1}((1-x_1+x_2)E_q+(1-x_1-x_2)k_0^z)(-\phi_{P_1}^P(x_1,b_0)\nonumber \\
	&\times \phi_{V_2}^s(x_2,b_2)
		+\phi_{P_1}^T(x_1,b_0)\phi_{V_2}^t(x_2,b_2))
		+r_{P_1}((1-x_1-x_2)E_q+(1-x_1+x_2)k_0^z)(\phi_{P_1}^T(x_1,b_0)\phi_{V_2}^s(x_2,b_2)\nonumber \\
	&-\phi_{P_1}^P(x_1,b_0)\phi_{V_2}^t(x_2,b_2))]
		h_{ne}(x_0,x_2,1-x_1,b_0,b_2)\exp[-S_B(t_{ne})-S_{P_1}(t_{ne})|_{b_1\to b_0}-S_{V_2}(t_{ne})]\nonumber\\
	&+\alpha_s(t_{ne}^\prime)[(1-x_2)(E_q-k_0^z)\phi_{P_1}^A(x_1,b_0)(\phi_{V_2}^s(x_2,b_2)+\phi_{V_2}^t(x_2,b_2))
		+r_{P_1}((2-x_1-x_2)E_q-(x_1-x_2)k_0^z)\nonumber \\
	&\times (\phi_{P_1}^P(x_1,b_0)\phi_{V_2}^s(x_2,b_2)+\phi_{P_1}^T(x_1,b_0)\phi_{V_2}^t(x_2,b_2))
		+r_{P_1}((x_1-x_2)E_q-(2-x_1-x_2)k_0^z)\nonumber \\
	&\times (\phi_{P_1}^T(x_1,b_0)\phi_{V_2}^s(x_2,b_2)+\phi_{P_1}^P(x_1,b_0)\phi_{V_2}^t(x_2,b_2))]
		\exp[-S_B(t_{ne}^\prime)-S_{P_1}(t_{ne}^\prime)|_{b_1\to b_0}-S_{V_2}(t_{ne}^\prime)]\nonumber \\
	&\times h_{ne}(x_0,1-x_2,1-x_1,b_0,b_2)\},
\end{align}
\begin{align} \label{eq:mep}
	\mathcal{M}_{e,P_1 V_2}^P=&16\pi^2C_Fm_B^2f_B
		\int k_{0\perp}dk_{0\perp} \int dx_0 dx_1 dx_2 \int b_0db_0 b_2db_2
		G_B(x_0,b_0,k_{0\perp}) (-\phi_{V_2}(x_2,b_2))\{\alpha_s(t_{ne}) \nonumber \\
	&\times [((1-x_1+x_2)E_q+(1-x_1-x_2)k_0^z)\phi_{P_1}^A(x_1,b_0)
		-r_{P_1}(1-x_1)(E_q-k_0^z)(\phi_{P_1}^P(x_1,b_0)-\phi_{P_1}^T(x_1,b_0))]\nonumber \\
	&\times h_{ne}(x_0,x_2,1-x_1,b_0,b_2)\exp[-S_B(t_{ne})-S_{P_1}(t_{ne})|_{b_1\to b_0}-S_{V_2}(t_{ne})]
		+\alpha_s(t_{ne}^\prime)[-(1-x_2)(E_q-k_0^z) \nonumber \\
	&\times \phi_{P_1}^A(x_1,b_0)
		+r_{P_1}(1-x_1)(E_q+k_0^z)(\phi_{P_1}^P(x_1,b_0)+\phi_{P_1}^T(x_1,b_0))]
		h_{ne}(x_0,1-x_2,1-x_1,b_0,b_2)\nonumber \\
	&\times \exp[-S_B(t_{ne}^\prime)-S_{P_1}(t_{ne}^\prime)|_{b_1\to b_0}-S_{V_2}(t_{ne}^\prime)]\},
\end{align}
\begin{align} \label{eq:fa}
	F_{a,P_1 V_2}=&8\pi C_Fm_B^2f_B
		\int dx_1 dx_2 \int b_1db_1 b_2db_2
		2\{\alpha_s(t_a)
		[x_2\phi_{P_1}^A(x_1,b_1)\phi_{V_2}(x_2,b_2)+2r_{P_1}r_{V_2}(1+x_2)\phi_{P_1}^P(x_1,b_1) \nonumber\\
	&\times \phi_{V_2}^s(x_2,b_2)
		-2r_{P_1}r_{V_2}(1-x_2)\phi_{P_1}^P(x_1,b_1)\phi_{V_2}^t(x_2,b_2)]
		h_a(1-x_1,x_2,b_1,b_2)S_t(x_2)\exp[-S_{P_1}(t_a)-S_{V_2}(t_a)]
		 \nonumber\\
	&+\alpha_s(t_a^\prime)[-(1-x_1)\phi_{P_1}^A(x_1,b_1)\phi_{V_2}(x_2,b_2)
		-2r_{P_1}r_{V_2}(2-x_1)\phi_{P_1}^P(x_1,b_1)\phi_{V_2}^s(x_2,b_2)
		-2r_{P_1}r_{V_2}x_1\phi_{P_1}^T(x_1,b_1) \nonumber\\
	&\times \phi_{V_2}^s(x_2,b_2)] h_a(x_2,1-x_1,b_2,b_1)S_t(x_1)\exp[-S_{P_1}(t_a^\prime)-S_{V_2}(t_a^\prime)]\},
\end{align}
\begin{align} \label{eq:fap}
	F_{a,P_1 V_2}^P=&8\pi C_Fm_B^2f_B\chi_B
		\int dx_1 dx_2 \int b_1db_1 b_2db_2
		4\{\alpha_s(t_a)
		[r_{V_2}x_2\phi_{P_1}^A(x_1,b_1)(\phi_{V_2}^s(x_2,b_2)-\phi_{V_2}^t(x_2,b_2))
		+2r_{P_1}\nonumber\\
	&\times \phi_{P_1}^P(x_1,b_1)\phi_{V_2}(x_2,b_2)]
		h_a(1-x_1,x_2,b_1,b_2) S_t(x_2) \exp[-S_{P_1}(t_a)-S_{V_2}(t_a)]
		+\alpha_s(t_a^\prime) [2r_{V_2}\phi_{P_1}^A(x_1,b_1) \nonumber \\
	&\times \phi_{V_2}^s(x_2,b_2)
		+r_{P_1}(1-x_1)(\phi_{P_1}^P(x_1,b_1)+\phi_{P_1}^T(x_1,b_1))\phi_{V_2}(x_2,b_2)]
		h_a(x_2,1-x_1,b_2,b_1)S_t(x_1) \nonumber \\
	&\times \exp[-S_{P_1}(t_a^\prime)-S_{V_2}(t_a^\prime)]\},
\end{align}
where
\begin{eqnarray} \label{chiB}
	\chi_B &&= \pi \int dk_{\perp}k_{\perp}\int_{x^d}^{x^u}dx \left(\frac{m_B}{2}+\frac{|\vec{k}_{\perp}|^2}{2x^2m_B}\right) K(\vec{k}) \left[(E_q+m_q)(E_Q+m_Q)+(E_q^2-m_q^2)\right],
\end{eqnarray}	
\begin{align} \label{eq:ma}
	\mathcal{M}_{a,P_1 V_2}=&16\pi^2C_Fm_B^2f_B
		\int k_{0\perp}dk_{0\perp} \int dx_0 dx_1 dx_2 \int b_0db_0 b_1db_1
		G_B(x_0,b_0,k_{0\perp})
		\{\alpha_s(t_{na}) [(1-x_1)(E_q+k_0^z)\phi_{P_1}^A(x_1,b_1) \nonumber \\
	&\times \phi_{V_2}(x_2,b_1)
		+r_{P_1}r_{V_2}((3-x_1+x_2)E_q+(1-x_1-x_2)k_0^z)\phi_{P_1}^P(x_1,b_1)\phi_{V_2}^s(x_2,b_1)
		+r_{P_1}r_{V_2}((1-x_1-x_2)E_q \nonumber \\
	&-(1+x_1-x_2)k_0^z)\phi_{P_1}^P(x_1,b_1)\phi_{V_2}^t(x_2,b_1)
		-r_{P_1}r_{V_2}((1-x_1-x_2)E_q+(3-x_1+x_2)k_0^z)\phi_{P_1}^T(x_1,b_1) \nonumber \\
	&\times \phi_{V_2}^s(x_2,b_1)
		+r_{P_1}r_{V_2}((1+x_1-x_2)E_q-(1-x_1-x_2)k_0^z)\phi_{P_1}^T(x_1,b_1)\phi_{V_2}^t(x_2,b_1)]
		h_{na}^1(1-x_1,x_2,b_0,b_1) \nonumber \\
	&\times \exp[-S_B(t_{na})-S_{P_1}(t_{na})-S_{V_2}(t_{na})|_{b_2\to b_1}]
		+\alpha_s(t_{na}^\prime) [-x_2(E_q-k_0^z)\phi_{P_1}^A(x_1,b_1)\phi_{V_2}(x_2,b_1) \nonumber \\
	&-r_{P_1}r_{V_2}((1-x_1+x_2)E_q+(1-x_1-x_2)k_0^z)(\phi_{P_1}^P(x_1,b_1)\phi_{V_2}^s(x_2,b_1)-\phi_{P_1}^T(x_1,b_1)\phi_{V_2}^t(x_2,b_1)) \nonumber \\
	&+r_{P_1}r_{V_2}((1-x_1-x_2)E_q+(1-x_1+x_2)k_0^z)(\phi_{P_1}^P(x_1,b_1)\phi_{V_2}^t(x_2,b_1)-\phi_{P_1}^T(x_1,b_1)\phi_{V_2}^s(x_2,b_1))] \nonumber \\
	&\times h_{na}^2(1-x_1,x_2,b_0,b_1)\exp[-S_B(t_{na}^\prime)-S_{P_1}(t_{na}^\prime)-S_{V_2}(t_{na}^\prime)|_{b_2\to b_1}]\},
\end{align}
\begin{align} \label{eq:mar}
	\mathcal{M}_{a,P_1 V_2}^R=&-16\pi^2C_Fm_B^2f_B
		\int k_{0\perp}dk_{0\perp} \int dx_0dx_1dx_2 \int b_0db_0b_1db_1 G_B(x_0,b_0,k_{0\perp})
		\{\alpha_s(t_{na}) [r_{V_2}((2-x_2)E_q+x_2k_0^z) \nonumber \\
	&\times \phi_{P_1}^A(x_1,b_1)(\phi_{V_2}^s(x_2,b_1)+\phi_{V_2}^t(x_2,b_1))
		-r_{P_1}((1+x_1)E_q-(1-x_1)k_0^z) (\phi_{P_1}^P(x_1,b_1)-\phi_{P_1}^T(x_1,b_1)) \nonumber \\
	&\times \phi_{V_2}(x_2,b_1)] h_{na}^1(1-x_1,x_2,b_0,b_1)
		\exp[-S_B(t_{na})-S_{P_1}(t_{na})-S_{V_2}(t_{na})|_{b_2\to b_1}]
		+\alpha_s(t_{na}^\prime) [r_{V_2}x_2 \nonumber \\
	&\times (E_q+k_0^z)\phi_{P_1}^A(x_1,b_1)(\phi_{V_2}^s(x_2,b_1)+\phi_{V_2}^t(x_2,b_1))
		-r_{P_1}(1-x_1)(E_q-k_0^z)(\phi_{P_1}^P(x_1,b_1)-\phi_{P_1}^T(x_1,b_1)) \nonumber \\
	&\times \phi_{V_2}(x_2,b_1)]h_{na}^2(1-x_1,x_2,b_0,b_1)
		\exp[-S_B(t_{na}^\prime)-S_{P_1}(t_{na}^\prime)-S_{V_2}(t_{na}^\prime)|_{b_2\to b_1}]\},
\end{align}
\begin{align} \label{eq:map}
	\mathcal{M}_{a,P_1 V_2}^P=&-16\pi^2C_Fm_B^2f_B
		\int k_{0\perp}dk_{0\perp} \int dx_0 dx_1 dx_2 \int b_0db_0 b_1db_1 G_B(x_0,b_0,k_{0\perp})
		\{\alpha_s(t_{na})[x_2(E_q-k_0^z) \nonumber \\
	&\times \phi_{P_1}^A(x_1,b_1)\phi_{V_2}(x_2,b_1)
		+r_{P_1}r_{V_2}((3-x_1+x_2)E_q+(1-x_1-x_2)k_0^z)\phi_{P_1}^P(x_1,b_1)\phi_{V_2}^s(x_2,b_1)
		 \nonumber \\
	&-r_{P_1}r_{V_2}((1-x_1-x_2)E_q+(3-x_1+x_2)k_0^z)\phi_{P_1}^P(x_1,b_1)\phi_{V_2}^t(x_2,b_1)
		+r_{P_1}r_{V_2}((1-x_1-x_2)E_q \nonumber \\
	&-(1+x_1-x_2)k_0^z)\phi_{P_1}^T(x_1,b_1)\phi_{V_2}^s(x_2,b_1)
		+r_{P_1}r_{V_2}((1+x_1-x_2)E_q-(1-x_1-x_2)k_0^z) \phi_{P_1}^T(x_1,b_1) \nonumber \\
	&\times \phi_{V_2}^t(x_2,b_1)]
		h_{na}^1(1-x_1,x_2,b_0,b_1) \exp[-S_B(t_{na})-S_{P_1}(t_{na})-S_{V_2}(t_{na})|_{b_2\to b_1}]
		+\alpha_s(t_{na}^\prime)[-(1-x_1)\nonumber \\
	&\times (E_q+k_0^z)\phi_{P_1}^A(x_1,b_1)\phi_{V_2}(x_2,b_1)
		-r_{P_1}r_{V_2}((1-x_1+x_2)E_q+(1-x_1-x_2)k_0^z) (\phi_{P_1}^P(x_1,b_1)\phi_{V_2}^s(x_2,b_1) \nonumber \\
	&-\phi_{P_1}^T(x_1,b_1)\phi_{V_2}^t(x_2,b_1))
		-r_{P_1}r_{V_2}((1-x_1-x_2)E_q+(1-x_1+x_2)k_0^z)(\phi_{P_1}^P(x_1,b_1)\phi_{V_2}^t(x_2,b_1)\nonumber \\
	&-\phi_{P_1}^T(x_1,b_1)\phi_{V_2}^s(x_2,b_1))]
		h_{na}^2(1-x_1,x_2,b_0,b_1) \exp[-S_B(t_{na}^\prime)-S_{P_1}(t_{na}^\prime)-S_{V_2}(t_{na}^\prime)|_{b_2\to b_1}]\}.
\end{align}
\begin{equation} \label{eq:ferfepfar}
    F_{e,P_1V_2}^R=F_{e,P_1V_2}, \quad F_{e,P_1V_2}^P=0, \quad F_{a,P_1V_2}^R = -F_{a,P_1V_2},
\end{equation}
In Eqs. (\ref{eq:me}) $\sim$ (\ref{eq:ferfepfar}), the amplitudes $F_{e,a}$ and $\mathcal{M}_{e,a}$ without superscript correspond to $(V-A)(V-A)$ operator insertion, while the amplitudes with superscripts $R$, and $P$ are relevant to $(V-A)(V+A)$ and $(S+P)(S-P)$ operator insertions, respectively.

The amplitude for the chromomagnetic penguin contribution is
\begin{align} \label{eq:mp}
	\mathcal{M}_{\mathrm{mp},P_1 V_2}=& 16\pi^2C_Fm_B^4f_B \left(\frac{C_F}{4\pi N_c}\right)
		\int k_{0\perp}dk_{0\perp} \int dx_0dx_1dx_2 \int b_0db_0b_1db_1b_2db_2 G_B(x_0,b_0,k_{0\perp})
		\{C_{8g}^{\mathrm{eff}}(t_g)\alpha_s^2(t_g) \nonumber \\
		&\times \bigl[x_1 [2(E_q-k_0^z)\phi_{P_1}^A(x_1,b_1)
			-r_{P_1}x_1(E_q+k_0^z)(\phi_{P_1}^P(x_1,b_1)+\phi_{P_1}^T(x_1,b_1))
			+2r_{P_1}((2E_q+k_0^z)\phi_{P_1}^P(x_1,b_1) \nonumber \\
		&-k_0^z\phi_{P_1}^T(x_1,b_1)) ] \phi_{V_2}(x_2,b_2)
			-r_{V_2} (1-x_2)((2-x_1)E_q-x_1k_0^z)\phi_{P_1}^A(x_1,b_1)(3\phi_{V_2}^s(x_2,b_2)+\phi_{V_2}^t(x_2,b_2))\nonumber \\
		&-r_{P_1} r_{V_2} x_1 (E_q-k_0^z) (\phi_{P_1}^P(x_1,b_1)+\phi_{P_1}^T(x_1,b_1))(3\phi_{V_2}^s(x_2,b_2)-\phi_{V_2}^t(x_2,b_2))
			+r_{P_1} r_{V_2} (1-x_2) ((1-2x_1)E_q \nonumber \\
		&-k_0^z)
			(\phi_{P_1}^P(x_1,b_1)-\phi_{P_1}^T(x_1,b_1))(3\phi_{V_2}^s(x_2,b_2)+\phi_{V_2}^t(x_2,b_2))\bigr]
			h_{g}(1-x_1,x_0(1-x_1),x_1(1-x_2),b_0,b_1,b_2) \nonumber \\
		&\times S_t(x_1) \exp[-S_B(t_g)-S_{P_1}(t_g)-S_{V_2}(t_g)]
			+C_{8g}^{\mathrm{eff}}(t_g^\prime)\alpha_s^2(t_g^\prime)2r_{P_1} (E_q+k_0^z)\phi_{P_1}^P(x_1,b_1)
			[2\phi_{V_2}(x_2,b_2) - r_{V_2} \nonumber \\
		&\times (1-x_2) (3\phi_{V_2}^s(x_2,b_2)+\phi_{V_2}^t(x_2,b_2))]
			h_g(x_0,x_0(1-x_1),1-x_2-x_0,b_1,b_0,b_2) S_t(x_0)\nonumber \\
		&\times \exp[-S_B(t_g^\prime)-S_{P_1}(t_g^\prime)-S_{V_2}(t_g^\prime)] \}.
\end{align}
The hard functions $h$, which arise from the Fourier transformation from momentum space to coordinate space, are defined in terms of the Bessel functions as follows
\begin{align}
	h_e(x_1,x_2,b_1,b_2)&=K_0(\sqrt{x_1x_2}m_Bb_1)[\theta(b_1-b_2) K_0(\sqrt{x_2}m_Bb_1)I_0(\sqrt{x_2}m_Bb_2) + (b_1 \leftrightarrow b_2)], \nonumber \\
	h_a(x_1,x_2,b_1,b_2)&=K_0(-i\sqrt{x_1x_2}m_Bb_1)[\theta(b_1-b_2) K_0(-i\sqrt{x_2}m_Bb_1)I_0(-i\sqrt{x_2}m_Bb_2)+ (b_1 \leftrightarrow b_2)], \nonumber \\
	h_{ne}(x_1,x_2,x_3,b_1,b_2)&=K_0(-i\sqrt{x_2x_3}m_Bb_2)[\theta(b_1-b_2) K_0(\sqrt{x_1x_3}m_Bb_1)I_0(\sqrt{x_1x_3}m_Bb_2)+ (b_1 \leftrightarrow b_2)], \nonumber \\	
	h_{na}^1(x_1,x_2,b_1,b_2)&=K_0(-i\sqrt{x_1x_2}m_Bb_1) [\theta(b_1-b_2) K_0(-i\sqrt{x_1x_2}m_Bb_1)I_0(-i\sqrt{x_1x_2}m_Bb_2) + (b_1 \leftrightarrow b_2)], \nonumber \\
	h_{na}^2(x_1,x_2,b_1,b_2)&=K_0(\sqrt{x_1+x_2-x_1x_2}m_Bb_1) [\theta(b_1-b_2) K_0(-i\sqrt{x_1x_2}m_Bb_1)I_0(-i\sqrt{x_1x_2}m_Bb_2) + (b_1 \leftrightarrow b_2)], \nonumber \\
	h_g(x_1, x_2, x_3, b_1, b_2, b_3) &= -K_0(\sqrt{x_2}m_Bb_1) K_0(-i\sqrt{x_3}m_Bb_3) \nonumber \\
	&\quad \times \int_{0}^{\pi/2} d\theta \tan\theta J_0(\sqrt{x_1}m_B\tan\theta b_1) J_0(\sqrt{x_1}m_B\tan\theta b_2) J_0(\sqrt{x_1}m_B\tan\theta b_3).
\end{align}
As we have explained, the scales $t$ are selected to the largest virtuality present in the diagrams, as detailed below
\begin{align}
	t_e=&\max(\sqrt{1-x_1}m_B, 1/b_0, 1/b_1), \quad
    t_e^\prime=\max(\sqrt{x_0}m_B, 1/b_0, 1/b_1), \nonumber \\
	t_a=&\max(\sqrt{x_2}m_B, 1/b_1, 1/b_2), \quad
	t_a^\prime=\max(\sqrt{1-x_1}m_B, 1/b_1, 1/b_2), \nonumber \\
	t_{ne}=&\max(\sqrt{x_0(1-x_1)}m_B, \sqrt{x_2(1-x_1)}m_B, 1/b_0, 1/b_2), \nonumber \\
    t_{ne}^\prime=&\max(\sqrt{x_0(1-x_1)}m_B, \sqrt{(1-x_2)(1-x_1)}m_B, 1/b_0, 1/b_2), \nonumber \\
	t_{na}=&\max(\sqrt{(1-x_1)x_2}m_B, 1/b_0, 1/b_1), \quad
    t_{na}^\prime=\max(\sqrt{(1-x_1)x_2}m_B, \sqrt{1-x_1+x_1x_2}m_B, 1/b_0, 1/b_1), \nonumber \\
	t_q		  	=&\max(\sqrt{1-x_1}m_B, \sqrt{x_1(1-x_2)}m_B, 1/b_0, 1/b_1), \quad
    t_q^\prime	=\max(\sqrt{x_0}m_B, \sqrt{|1-x_2-x_0|}m_B, 1/b_0, 1/b_1), \nonumber \\
	t_g			=&\max(\sqrt{1-x_1}m_B, \sqrt{x_1(1-x_2)}m_B, 1/b_0, 1/b_1, 1/b_2), \quad
    t_g^\prime	=\max(\sqrt{x_0}m_B, \sqrt{|1-x_2-x_0|}m_B, 1/b_0, 1/b_1, 1/b_2), \nonumber \\
	l_{q}^2			=& x_1(1-x_2)m_B^2, \quad
	l_{q}^{\prime 2}= (1-x_2-x_0)m_B^2,
\end{align}
}

As for the $B\to V_1P_2$ transitions, most of the amplitudes can be obtained by applying a substitution to those of the $B\to P_1V_2$ case.
We define the substitution $S$ as
\begin{align}
	S=\left\{\phi_{P_1}^{A,P,T}\to\phi_{V_1}^{\_,s,t}, \phi_{V_2}^{\_,s,t}\to\phi_{P_2}^{A,P,T},r_{P_1}\to -r_{V_1}, r_{V_2}\to r_{P_2}\right\}.
\end{align}
The resulting amplitudes are then
\begin{align}
	F_{e,V_1P_2} =& F_{e,P_1V_2}|_{S}, \quad F_{e,V_1 P_2}^R=-F_{e,V_1 P_2}, \quad
	\mathcal{M}_{e,V_1P_2}^{(R)} = \mathcal{M}_{e,P_1V_2}^{(R)}|_{S},\quad \mathcal{M}_{e,V_1P_2}^P = -\mathcal{M}_{e,P_1V_2}^P|_{S}, \nonumber \\
	F_{a,V_1P_2} =& F_{a,P_1V_2}|_{S}, \quad F_{a,V_1P_2}^R = -F_{a,V_1P_2}, \quad F_{a,V_1P_2}^P = -F_{a,P_1V_2}^P|_{S}, \nonumber \\
	\mathcal{M}_{a,V_1P_2}^{(P)} =& \mathcal{M}_{a,P_1V_2}^{(P)}|_{S}, \quad \mathcal{M}_{a,V_1P_2}^R = -\mathcal{M}_{a,P_1V_2}^R|_{S}, \quad
	\mathcal{M}_{\mathrm{ql},V_1P_2}^{(t)} = \mathcal{M}_{\mathrm{ql},P_1V_2}^{(t)}|_{S}, \quad
	\mathcal{M}_{\mathrm{mp},V_1P_2} = \mathcal{M}_{\mathrm{mp},P_1 V_2}|_{S}.
\end{align}
Nevertheless, there exists a special amplitude absent in the $B\to P_1V_2$ transition due to the meson's parity property.
Its explicit form is presented below
\begin{align}\label{eq:fep}
	F_{e,V_1 P_2}^P=&-16\pi^2C_Fm_B^2f_B\left(\frac{f_{P_2}}{2\sqrt{2N_c}}\right)
		\int k_{0\perp}dk_{0\perp} \int dx_0 dx_1 \int b_0db_0b_1db_1
		G_B(x_0,b_0,k_{0\perp}) 2r_{P_2} \{\alpha_s(t_e)
		[-(E_q-k_0^z) \nonumber \\
	&\times \phi_{V_1}(x_1,b_1)
		+r_{V_1} ((3-x_1)E_q+(1-x_1)k_0^z)\phi_{V_1}^s(x_1,b_1)
		+r_{V_1} ((1-x_1)E_q-(1+x_1)k_0^z)\phi_{V_1}^t(x_1,b_1)] S_t(x_1) \nonumber \\
	&\times	h_e(x_0,1-x_1,b_0,b_1)\exp[-S_B(t_e)-S_{V_1}(t_e)]
		+\alpha_s(t_e^\prime)2r_{V_1}(E_q+k_0^z)\phi_{V_1}^s(x_1,b_1)h_e(1-x_1,x_0,b_1,b_0) S_t(x_0) \nonumber \\
	&\times \exp[-S_B(t_e^\prime)-S_{V_1}(t_e^\prime)]\}.
\end{align}

\section{\uppercase{The Wilson coefficients for various diagrams} \label{app:wilson}}
For $B\to PV$ decays, the Wilson coefficients corresponding to various LO and NLO diagrams, along with the constants combined with the amplitudes, are summarized in Tables~\ref{tab:wilson1} and \ref{tab:wilson2}. The contributions of the color-octet components consist of five types of amplitudes, as indicated in Eq.~(\ref{eq:fm8}). The Wilson coefficients for $\mathcal{M}_{e}^{(P,R),8}$, $F_{a}^{(P,R),8}$, and $\mathcal{M}_{a}^{(P,R),8}$ are identical to those of the color-singlet contributions. For the cases of $F_{e}^{(P,R),8}$ and  $\mathcal{M}_{e}^{(P,R),8^\prime}$, we define the substitutions
$S_2=\left\{a_{1(2)}\to C_{1(2)}, a_{i(i+1)}\to C_{i+1(i)},\ \text{for i = 3,5,7,9} \right\}$ and $S_3=\left\{C_{j(j+1)}\to C_{j+1(k)},\ \text{for j = 1,3,5,7,9} \right\} $. The associated Wilson coefficients can be derived from those of the color-singlet cases through the substitutions $S_2$ and $S_3$, respectively.

\begin{table*}
  \caption{\label{tab:wilson1}
  Wilson coefficients corresponding to various topological diagrams in the decay amplitudes of
  $B \to \rho\pi,\ \rho\eta_{q,s},\ \omega\pi, \ \phi\pi, \ K^*\bar{K}$ and $\bar{K}^*K$.}
  \renewcommand\arraystretch{1.35}
  \begin{ruledtabular}
    \begin{tabular}{cccccccccccccc}
		& $\rho^-\pi^0$ & $\rho^+\pi^-$ & $\rho^0\pi^0$ & $\rho^-\eta_q$ & $\rho^0\eta_q$ & $\rho^-\eta_s$ & $\rho^0\eta_s$ & $K^{*-}K^0$ & $K^{*-}K^+$ & $\bar{K}^{*0}K^0$ \\
		& $\pi^-\rho^0$ & $\pi^+\rho^-$ & 				& $\pi^-\omega$  & $\pi^0\omega$  & $\pi^-\phi$    & $\pi^0\phi$ 	& $K^-K^{*0}$ & $K^-K^{*+}$ & $\bar{K}^0K^{*0}$ \\
		\hline
		$R$											& $\sqrt{2}$			& 1 			& $2$ 				 	& $\sqrt{2}$				& 2						& 1			& $\sqrt{2}$ & 1 			& 1 			& 1 			\\
		\multirow{2}{*}{$F_{e, M_1M_2}$}			& $a_2+a_3^u$ 			& $a_1+a_4^u$	& $-a_2-a_3^u$		 	& $a_2+a_3^u$				& $-a_2-a_3^u$			& $a_3^s$ 	& $-a_3^s$ 	 & $a_4^s$ 		& 0 			& $a_4^s$ 		\\
													& $\quad-a_3^d-a_4^d$ 	&				& $\quad+a_3^d+a_4^d$	& $\quad+a_3^d+a_4^d$		& $\quad-a_3^d-a_4^d$	& 			& 			 &				&				&				\\
		$F_{e, M_1M_2}^R$ 							& $a_5^u-a_5^d$ 		& 0				& $-a_5^u+a_5^d$		& $a_5^u+a_5^d$				& $-a_5^u-a_5^d$		& $a_5^s$ 	& $-a_5^s$ 	 & 0			& 0 			& 0 			\\
		$F_{e, M_1M_2}^P$ 							& $-a_6^d$				& $a_6^u$		& $a_6^d$				& $a_6^d$					& $-a_6^d$				& 0 		& 0 		 & $a_6^s$		& 0 			& $a_6^s$		\\
		\multirow{2}{*}{$\mathcal{M}_{e,M_1M_2}$ }  & $C_2+C_4^u$ 			& $C_1+C_3^u$	& $-C_2-C_4^u$			& $C_2+C_4^u$				& $-C_2-C_4^u$			& $C_4^s$ 	& $-C_4^s$ 	 & $C_3^s$		& 0 			& $C_3^s$		\\
													& $\quad-C_3^d-C_4^d$ 	&				& $\quad+C_3^d+C_4^d$	& $\quad+C_3^d+C_4^d$		& $\quad-C_3^d-C_4^d$	& 			& 			 &				&				&				\\
		$\mathcal{M}_{e,M_1M_2}^R$ 					& $-C_5^d$				& $C_5^u$		& $C_5^d$				& $C_5^d$					& $-C_5^d$				& 0 		& 0 		 & $C_5^s$		& 0 			& $C_5^s$		\\
		$\mathcal{M}_{e,M_1M_2}^P$ 					& $C_6^u-C_6^d$			& 0				& $-C_6^u+C_6^d$		& $C_6^u+C_6^d$				& $-C_6^u-C_6^d$		& $C_6^s$ 	& $-C_6^s$ 	 & 0			& 0 			& 0 			\\
		\multirow{2}{*}{$F_{e, M_2M_1}$}			& $a_1+a_4^u$			& 0 			& $-a_2-a_3^u$		 	& $a_1+a_4^u$ 				& $a_2+a_3^u$ 			& 0 		& 0 		 & 0 			& 0 			& 0 			\\
													&						&				& $\quad+a_3^d+a_4^d$	&							& $\quad-a_3^d-a_4^d$ 	& 			& 			 &   			&   			&   			\\
		$F_{e, M_2M_1}^R$ 							& 0						& 0 			& $-a_5^u+a_5^d$		& 0 						& $a_5^u-a_5^d$ 		& 0			& 0			 & 0 			& 0 			& 0				\\
		$F_{e, M_2M_1}^P$ 							& $a_6^u$				& 0 			& $a_6^d$				& $a_6^u$ 					& $-a_6^d$				& 0			& 0			 & 0 			& 0 			& 0 			\\
		\multirow{2}{*}{$\mathcal{M}_{e,M_2M_1}$}   & $C_1+C_3^u$			& 0 			& $-C_2-C_4^u$			& $C_1+C_3^u$ 				& $C_2+C_4^u$ 			& 0			& 0			 & 0 			& 0 			& 0 			\\
													&						&				& $\quad+C_3^d+C_4^d$	& 			 				& $\quad-C_3^d-C_4^d$ 	& 			& 			 & 			 	& 			 	& 			 	\\
		$\mathcal{M}_{e,M_2M_1}^R$ 					& $C_5^u$				& 0 			& $C_5^d$				& $C_5^u$ 					& $-C_5^d$				& 0			& 0			 & 0 			& 0 			& 0 			\\
		$\mathcal{M}_{e,M_2M_1}^P$ 					& 0						& 0 			& $-C_6^u+C_6^d$		& 0 						& $C_6^u-C_6^d$			& 0			& 0			 & 0 			& 0 			& 0				\\
		\multirow{2}{*}{$F_{a,M_1M_2}$}   			& $a_1+a_4^u$			& $a_2+a_3^u$	& $a_2+a_3^u$			& $a_1+a_4^u$ 				& $a_2+a_3^u$ 			& 0			& 0			 & 0 			& $a_3^s$ 		& $a_3^s$ 		\\
													&						&				& $\quad+a_3^d+a_4^d$	& 			 				& $\quad-a_3^d-a_4^d$ 	& 			& 			 & 			 	& 			 	& 			 	\\
		$F_{a,M_1M_2}^R$ 							& 0						& $a_5^u$		& $a_5^u+a_5^d$			& 0 						& $a_5^u-a_5^d$ 		& 0			& 0			 & 0 			& $a_5^s$ 		& $a_5^s$ 		\\
		$F_{a,M_1M_2}^P$ 							& $a_6^u$				& 0				& $a_6^d$				& $a_6^u$ 					& $-a_6^d$				& 0			& 0			 & 0 			& 0 			& 0 			\\
		\multirow{2}{*}{$\mathcal{M}_{a,M_1M_2}$}   & $C_1+C_3^u$			& $C_2+C_4^u$	& $C_2+C_4^u$			& $C_1+C_3^u$ 				& $C_2+C_4^u$ 			& 0			& 0			 & 0 			& $C_4^s$ 		& $C_4^s$ 		\\
													&						&				& $\quad+C_3^d+C_4^d$	& 			 				& $\quad-C_3^d-C_4^d$ 	& 			& 			 & 			 	& 			 	& 			 	\\
		$\mathcal{M}_{a,M_1M_2}^R$ 					& $C_5^u$				& 0				& $C_5^d$				& $C_5^u$ 					& $-C_5^d$				& 0			& 0			 & 0 			& 0 			& 0 			\\
		$\mathcal{M}_{a,M_1M_2}^P$ 					& 0						& $C_6^u$		& $C_6^u+C_6^d$			& 0 						& $C_6^u-C_6^d$			& 0			& 0			 & 0 			& $C_6^s$ 		& $C_6^s$ 		\\
		\multirow{2}{*}{$F_{a,M_2M_1}$}   			& $-a_1-a_4^u$			& $a_3^d+a_4^d$	& $a_2+a_3^u$			& $a_1+a_4^u$ 				& $a_2+a_3^u$ 			& 0			& 0			 & $a_1+a_4^u$	& $a_2+a_3^u$	& $a_3^d+a_4^d$	\\
													&						&				& $\quad+a_3^d+a_4^d$	& 			 				& $\quad-a_3^d-a_4^d$ 	& 			& 			 & 			 	& 			 	& 			 	\\
		$F_{a,M_2M_1}^R$ 							& 0						& $a_5^d$		& $a_5^u+a_5^d$			& 0 						& $a_5^u-a_5^d$ 		& 0			& 0			 & 0			& $a_5^u$		& $a_5^d$		\\
		$F_{a,M_2M_1}^P$ 							& $-a_6^u$				& $a_6^d$		& $a_6^d$				& $a_6^u$ 					& $-a_6^d$				& 0			& 0			 & $a_6^u$		& 0				& $a_6^d$		\\
		\multirow{2}{*}{$\mathcal{M}_{a,M_2M_1}$}   & $-C_1-C_3^u$			& $C_3^d+C_4^d$	& $C_2+C_4^u$			& $C_1+C_3^u$ 				& $C_2+C_4^u$ 			& 0			& 0			 & $C_1+C_3^u$	& $C_2+C_4^u$	& $C_3^d+C_4^d$	\\
													&						&				& $\quad+C_3^d+C_4^d$	& 							& $\quad-C_3^d-C_4^d$ 	& 			& 			 & 			 	& 			 	& 			 	\\
		$\mathcal{M}_{a,M_2M_1}^R$ 					& $-C_5^u$				& $C_5^d$		& $C_5^d$				& $C_5^u$ 					& $-C_5^d$				& 0			& 0			 & $C_5^u$		& 0				& $C_5^d$		\\
		$\mathcal{M}_{a,M_2M_1}^P$ 					& 0						& $C_6^d$		& $C_6^u+C_6^d$			& 0 						& $C_6^u-C_6^d$			& 0			& 0			 & 0 			& $C_6^u$		& $C_6^d$		\\
		$\mathcal{M}_{\mathrm{ql},M_1M_2}^{(u,c,t)}$& -1 					& 1 			& 1 					& 1							& -1					& 0			& 0			 & 0 			& 0				& 1				\\
		$\mathcal{M}_{\mathrm{mp},M_1M_2}$ 			& -1 					& 1 			& 1 					& 1							& -1					& 0			& 0			 & 1			& 0				& 1				\\
		$\mathcal{M}_{\mathrm{ql},M_2M_1}^{(u,c,t)}$& 1 					& 0 			& 1 					& 1							& -1					& 0			& 0			 & 0			& 0				& 0				\\
		$\mathcal{M}_{\mathrm{mp},M_2M_1}$ 			& 1 					& 0 			& 1 					& 1							& -1					& 0			& 0			 & 0			& 0				& 0				\\
	\end{tabular}
  \end{ruledtabular}
\end{table*}

\begin{table*}
  \caption{\label{tab:wilson2}
  Wilson coefficients corresponding to various topological diagrams in the decay amplitudes of
  $B \to K^*\pi,\ \rho K,\ K^*\eta_{q,s},\ \omega K$ and $\phi K$.}
  \renewcommand\arraystretch{1.35}
  \begin{ruledtabular}
    \begin{tabular}{cccccccccccccc}
		& $\bar{K}^{*0}\pi^-$ & $K^{*-}\pi^0$ & $K^{*-}\pi^+$ & $\bar{K}^{*0}\pi^0$ & $K^{*-}\eta_q$ & $\bar{K}^{*0}\eta_q$ & $K^{*-}\eta_s$ & $\bar{K}^{*0}\eta_s$   \\
		& $\bar{K}^0\rho^-$   & $K^-\rho^0$   & $K^-\rho^+$   & $\bar{K}^0\rho^0$   & $K^-\omega$    & $\bar{K}^0\omega$    & $K^-\phi$ 	 & $\bar{K}^0\phi$ 		  \\
		\hline
		$R$											& 1				& $\sqrt{2}$		& 1				& $\sqrt{2}$		& $\sqrt{2}$	 	& $\sqrt{2}$	 	& 1 			& 1 			\\
		$F_{e, M_1M_2}$								& 0 			& $a_2+a_3^u-a_3^d$ & 0 			& $a_2+a_3^u-a_3^d$	& $a_2+a_3^u+a_3^d$ & $a_2+a_3^u+a_3^d$ & $a_3^s+a_4^s$ & $a_3^s+a_4^s$ \\
		$F_{e, M_1M_2}^R$ 							& 0 			& $a_5^u-a_5^d$ 	& 0 			& $a_5^u-a_5^d$		& $a_5^u+a_5^d$ 	& $a_5^u+a_5^d$  	& $a_5^s$		& $a_5^s$		\\
		$F_{e, M_1M_2}^P$ 							& 0 			& 0 				& 0 			& 0					& 0 			 	& 0 			 	& $a_6^s$		& $a_6^s$		\\
		$\mathcal{M}_{e,M_1M_2}$					& 0 			& $C_2+C_4^u-C_4^d$	& 0 			& $C_2+C_4^u-C_4^d$	& $C_2+C_4^u+C_4^d$ & $C_2+C_4^u+C_4^d$ & $C_3^s+C_4^s$ & $C_3^s+C_4^s$ \\
		$\mathcal{M}_{e,M_1M_2}^R$ 					& 0 			& 0 				& 0 			& 0					& 0 			 	& 0 			 	& $C_5^s$		& $C_5^s$		\\
		$\mathcal{M}_{e,M_1M_2}^P$ 					& 0 			& $C_6^u-C_6^d$ 	& 0 			& $C_6^u-C_6^d$		& $C_6^u+C_6^d$ 	& $C_6^u+C_6^d$  	& $C_6^s$		& $C_6^s$		\\
		$F_{e, M_2M_1}$		   						& $a_4^d$ 		& $a_1+a_4^u$ 		& $a_1+a_4^u$ 	& $-a_4^d$			& $a_1+a_4^u$ 	 	& $a_4^d$ 		 	& 0			 	& 0			 	\\
		$F_{e, M_2M_1}^R$ 							& 0 			& 0 				& 0 			& 0					& 0 			 	& 0			 	 	& 0			 	& 0			 	\\
		$F_{e, M_2M_1}^P$ 							& $a_6^d$ 		& $a_6^u$ 			& $a_6^u$ 		& $-a_6^d$			& $a_6^u$ 		 	& $a_6^d$		 	& 0			 	& 0			 	\\
		$\mathcal{M}_{e,M_2M_1}$					& $C_3^d$ 		& $C_1+C_3^u$ 		& $C_1+C_3^u$ 	& $-C_3^d$			& $C_1+C_3^u$ 	 	& $C_3^d$		 	& 0			 	& 0			 	\\
		$\mathcal{M}_{e,M_2M_1}^R$ 					& $C_5^d$ 		& $C_5^u$ 			& $C_5^u$ 		& $-C_5^d$			& $C_5^u$ 		 	& $C_5^d$		 	& 0			 	& 0			 	\\
		$\mathcal{M}_{e,M_2M_1}^P$ 					& 0 			& 0 				& 0 			& 0					& 0 			 	& 0			 	 	& 0			 	& 0			 	\\
		$F_{a,M_1M_2}$								& $a_1+a_4^u$ 	& $a_1+a_4^u$ 		& $a_4^d$ 		& $-a_4^d$			& $a_1+a_4^u$ 	 	& $a_4^d$ 		 	& 0			 	& 0			 	\\
		$F_{a,M_1M_2}^R$ 							& 0 			& 0 				& 0 			& 0					& 0 			 	& 0			 	 	& 0			 	& 0			 	\\
		$F_{a,M_1M_2}^P$ 							& $a_6^u$ 		& $a_6^u$ 			& $a_6^d$ 		& $-a_6^d$			& $a_6^u$ 		 	& $a_6^d$		 	& 0			 	& 0			 	\\
		$\mathcal{M}_{a,M_1M_2}$					& $C_1+C_3^u$ 	& $C_1+C_3^u$ 		& $C_3^d$ 		& $-C_3^d$			& $C_1+C_3^u$ 	 	& $C_3^d$		 	& 0			 	& 0			 	\\
		$\mathcal{M}_{a,M_1M_2}^R$ 					& $C_5^u$ 		& $C_5^u$ 			& $C_5^d$ 		& $-C_5^d$			& $C_5^u$ 		 	& $C_5^d$		 	& 0			 	& 0			 	\\
		$\mathcal{M}_{a,M_1M_2}^P$ 					& 0 			& 0 				& 0 			& 0					& 0 			 	& 0			 	 	& 0			 	& 0			 	\\
		$F_{a,M_2M_1}$								& 0 			& 0 				& 0 			& 0					& 0			 		& 0			 	 	& $a_1+a_4^u$ 	& $a_4^d$ 		\\
		$F_{a,M_2M_1}^R$ 							& 0 			& 0 				& 0 			& 0					& 0			 		& 0			 	 	& 0 			& 0			 	\\
		$F_{a,M_2M_1}^P$ 							& 0 			& 0 				& 0 			& 0					& 0			 		& 0			 	 	& $a_6^u$ 		& $a_6^d$		\\
		$\mathcal{M}_{a,M_2M_1}$					& 0 			& 0 				& 0 			& 0					& 0			 		& 0			 	 	& $C_1+C_3^u$ 	& $C_3^d$		\\
		$\mathcal{M}_{a,M_2M_1}^R$ 					& 0 			& 0 				& 0 			& 0					& 0			 		& 0			 	 	& $C_5^u$ 		& $C_5^d$		\\
		$\mathcal{M}_{a,M_2M_1}^P$ 					& 0 			& 0 				& 0 			& 0					& 0			 		& 0			 	 	& 0			 	& 0			 	\\
		$\mathcal{M}_{\mathrm{ql},M_1M_2}^{(u,c,t)}$& 0 			& 0 				& 0 			& 0					& 0			 		& 0			 	 	& 1 			& 1			 	\\
		$\mathcal{M}_{\mathrm{mp},M_1M_2}$ 			& 0 			& 0 				& 0 			& 0					& 0			 		& 0			 	 	& 1 			& 1			 	\\
		$\mathcal{M}_{\mathrm{ql},M_2M_1}^{(u,c,t)}$& 1 			& 1 				& 1 			& -1				& 1			 		& 1			 	 	& 0 			& 0			 	\\
		$\mathcal{M}_{\mathrm{mp},M_2M_1}$ 			& 1 			& 1 				& 1 			& -1				& 1			 		& 1			 	 	& 0 			& 0			 	\\
	\end{tabular}
  \end{ruledtabular}
\end{table*}

\end{widetext}

\end{document}